%% file: manuscript.tex
\documentclass[a4paper,11pt]{article}
\usepackage{jheppub}
\usepackage{amsmath,amssymb,mathtools}
\usepackage{booktabs}
\usepackage{float}
\usepackage{flafter}
\usepackage{array}
\usepackage{multirow}
\usepackage{nicematrix}
\usepackage{tikz}
\usepackage{CJKutf8}
\usetikzlibrary{arrows.meta,calc,fadings,fit,positioning,shapes.geometric}

\definecolor{mutblue}{HTML}{DCEAF7}
\definecolor{periodone}{HTML}{E7F2E4}
\definecolor{periodtwo}{HTML}{FBE9D8}
\definecolor{orbitone}{HTML}{F6D6D6}
\definecolor{orbittwo}{HTML}{D8E8F7}
\definecolor{orbitthree}{HTML}{E5DCF4}
\definecolor{orbitfour}{HTML}{D9EFE4}
\definecolor{orbitfive}{HTML}{F8E8BF}
\definecolor{accentblue}{HTML}{3F6F9F}
\definecolor{accentorange}{HTML}{C87936}
\definecolor{accentgreen}{HTML}{4F8A62}

\tikzset{
  mutable/.style={draw=black!70,rounded corners=1.2pt,fill=mutblue,
    minimum width=11mm,minimum height=7mm,inner sep=2pt,font=\scriptsize},
  frozenone/.style={draw=accentgreen!75!black,rounded corners=1.2pt,
    fill=periodone,minimum width=11mm,minimum height=7mm,
    inner sep=2pt,font=\scriptsize},
  frozentwo/.style={draw=accentorange!80!black,rounded corners=1.2pt,
    fill=periodtwo,minimum width=11mm,minimum height=7mm,
    inner sep=2pt,font=\scriptsize},
  qarrow/.style={-{Stealth[length=2.1mm,width=1.45mm]},semithick,black!75},
  flowbox/.style={draw=black!55,rounded corners=2pt,fill=black!2,
    align=center,minimum height=10mm,text width=28mm,font=\small},
  flowarrow/.style={-{Stealth[length=2.5mm]},thick,accentblue}
}

\newcommand{\Gr}{\operatorname{Gr}}
\newcommand{\br}[1]{\langle #1\rangle}
\newcommand{\cP}{\mathcal P}

\title{Form Factor Alphabets and Antipodal Duality from Cluster Algebras}

\author[a,b,c]{Song He (何颂)}
\author[a,c]{Jiahao Liu (刘家昊)}

\affiliation[a]{New Cornerstone Laboratory, Institute of Theoretical Physics,
Chinese Academy of Sciences, Beijing 100190, China}
\affiliation[b]{School of Fundamental Physics and Mathematical Sciences,
Hangzhou Institute for Advanced Study, UCAS \& ICTP-AP,
Hangzhou 310024, China}
\affiliation[c]{School of Physical Sciences, University of Chinese Academy of
Sciences, No.~19A Yuquan Road, Beijing 100049, China}

\emailAdd{songhe@itp.ac.cn}
\emailAdd{liujiahao@itp.ac.cn}

\abstract{
We develop a cluster-algebraic description of symbol alphabets for chiral
stress-tensor form factors in planar $\mathcal N=4$ super-Yang--Mills theory.
Motivated by unfolding the three-point $C_2$ quiver, we propose a periodic
framework for $n$-point form-factor alphabets based on folding a
$\Gr(4,2n)$ quiver with two periods of frozen nodes.  At four points, the
folding leads to the $\widetilde{\mathrm{VI}}$ cluster algebra, which has
infinitely many cluster variables.  Tropical truncation selects a finite set
of rational coordinates and four physical limit rays.  The
$A_1^{(1)}$ mutation sequences approaching these rays generate the four
square roots and the corresponding algebraic-letter spaces.  Within the
resulting finite set of rational candidates, antipodal closure excludes
precisely eight additional letters.  On the parity-preserving surface,
the antipodal map is realized, after a further folding, by two commuting
mutations followed by a relabelling.  Furthermore,
we find an alternative quiver within the same mutation class that becomes
self-antipodal after parity folding, mirroring the antipodal self-duality of
the four-point MHV form factor.  Its double- and triple-collinear boundaries
reproduce, respectively, the $C_2$ algebra of the three-point form factor and
the $A_3$ algebra of the six-point amplitude; after folding the latter
to $C_2$, the antipodal map exchanges the two $C_2$
boundaries.
}
\keywords{planar \(\mathcal N=4\) super-Yang-Mills theory, stress-tensor form factors, cluster algebras, periodic momentum twistors, symbol alphabets, antipodal duality}

\begin{document}
\begin{CJK*}{UTF8}{gbsn}
\maketitle
\end{CJK*}
\raggedbottom

\section{Introduction}
\label{sec:introduction}

Scattering amplitudes and form factors are key physical quantities in gauge
theory: amplitudes describe transitions between on-shell states, whereas form
factors are matrix elements of local gauge-invariant operators between the
vacuum and on-shell multiparticle states.  Perturbative results for both
quantities in planar $\mathcal N=4$ super-Yang--Mills (SYM) theory are available
to unusually high loop order and provide an important testing ground for bootstrap
methods~\cite{Dixon:2011pw,Drummond:2014ffa,Dixon:2016nkn,
Drummond:2017ssj,Drummond:2018caf,Caron-Huot:2019vjl,
Dixon:2020bbt,Dixon:2022xqh,Dixon:2022rse,Dixon:2023kop,
He:2025tyv,He:2026ddp}.  Many such multi-loop amplitudes and form factors can be
expressed in terms of multiple polylogarithms, whose symbols are built from
logarithmic arguments drawn from a finite
alphabet~\cite{Goncharov:2010jf,Duhr:2011zq,Duhr:2012fh}.  A candidate alphabet
constrains the possible singular loci and is a basic input to the bootstrap.
Figure~\ref{fig:amplitudes-form-factors-roadmap}
provides an overview of the amplitude and form factor alphabets studied below
and their kinematic relations.

\input{figures/amplitudes_form_factors_roadmap.tex}

The kinematics of planar amplitudes admits a direct geometric description.
Ordered momenta define a closed null polygon in dual space.  The dual points are
represented by lines in momentum twistor space, on which the dual conformal
group acts linearly~\cite{Hodges:2009hk}.
The momentum twistor four-brackets of a planar $n$-point amplitude are
Pl\"ucker coordinates on $\Gr(4,n)$, whose homogeneous coordinate ring
carries a cluster algebra structure~\cite{Fomin:2001ap,Scott:2003vk}.  The
9-letter hexagon and 42-letter heptagon alphabets are organized by the finite
cluster algebras $\Gr(4,6)\simeq A_3$ and $\Gr(4,7)\simeq E_6$
\cite{Golden:2013xva,Golden:2014xqa}.  The so-called cluster adjacency further
requires consecutive symbol letters to appear together in a common
cluster~\cite{Drummond:2017ssj,Drummond:2018dfd}.  The cluster algebra associated
with $\Gr(4,8)$ is of infinite type, whereas fixed-loop eight-point amplitudes
involve finite alphabets.  This is seen in the two-loop NMHV amplitudes at
general multiplicity and in the three-loop MHV
octagon~\cite{He:2019jee,He:2020vob,Li:2021bwg}.  Finite subalgebras and
tropical fans have been used to obtain finite rational alphabets, while limit
rays of infinite mutation sequences give rise to square roots and algebraic
letters~\cite{Drummond:2019qjk,Drummond:2019cxm,Henke:2019hve,
Chicherin:2020umh,He:2020lcu,He:2021non,Henke:2021ity,He:2021esx,
He:2021eec}.  These developments separate the ambient cluster structure from
the finite alphabet selected by a physical problem.
For example, geometric analyses of intersections in momentum-twistor space
provide a complementary way to construct symbol letters and relate them
to positivity and Landau
singularities~\cite{Yang:2022gko,He:2022tph,He:2023qld,He:2024lba}.
Related cluster structures on partial flag varieties have begun to organize
singularities and rational and algebraic letters beyond dual-conformal
kinematics~\cite{Pokraka:2025ali,Bossinger:2025rhf}.

Form-factor kinematics can be described using periodic momentum twistors.  For
an $n$-point form factor, the sum of the external momenta is the operator
momentum $q$, so the dual contour is translated by $q$ after $n$ null edges
rather than closing.  The periodic
Wilson-loop picture first appeared at one loop~\cite{Brandhuber:2010ad}, and
momentum twistor geometry was subsequently developed for stress-tensor form
factors~\cite{Bork:2014eqa}.  These descriptions have been used extensively in
recent calculations~\cite{Guo:2021bym,Guo:2022qgv,Li:2024qtn}.  Despite this
natural periodic geometry, a systematic relation between
periodic Grassmannian quivers and form-factor alphabets, analogous to the
Grassmannian construction for amplitudes, has remained unclear.

At three points, the cluster construction is especially simple: unfolding the
$C_2$ quiver associated with the form factor alphabet yields a $\Gr(4,6)$ quiver with two
periods of frozen nodes.  We generalize this construction by mutating a
$\Gr(4,2n)$ quiver with two periods of frozen nodes to a $\mathbb Z_2$-symmetric
seed, then folding it by identifying nodes that represent the same function on
periodic kinematics.  Two questions then arise: can periodic momentum twistor
coordinates realize the folded exchange relations, and, when the resulting
cluster algebra is infinite, how is a finite physical alphabet selected?  We
answer these questions explicitly at four points.

At four points, we fold a $\Gr(4,8)$ quiver with two periods of frozen nodes
to obtain a rank-five cluster algebra whose principal quiver is of
$\widetilde{\mathrm{VI}}$ mutation type.  An auxiliary square-root redefinition and
rephasing bring its folded exchange relations to standard
skew-symmetrizable form.  Although its mutation class is finite, the
resulting cluster algebra has infinitely many variables.  An initial tropical
truncation retains precisely $64$ non-frozen rational letters and two physical
limit rays.  Refining the tropical input recovers the other two physical
limit rays while introducing eight additional rational coordinates.  The
$A_1^{(1)}$ mutation sequences approaching the four rays generate the four
physical square roots and five associated algebraic letters for each, giving
$20$ algebraic letters in total.  Requiring antipodal closure excludes the
eight additional coordinates.  Together, the remaining letters form the
physical alphabet
\[
  88=64_{\mathrm{rat}}+4_{\mathrm{roots}}+4\times5_{\mathrm{alg}},
\]
which is compatible with the four-point MHV and NMHV bootstrap results through
four and three loops, respectively~\cite{Dixon:2022rse,DixonLi:toappear,
He:2026ddp,HeLiuYang:toappear}.

The antipodal closure used above is motivated by known antipodal dualities.
On the parity-preserving surface, the six-point MHV amplitude is related to
the three-point stress-tensor form factor by the antipode operation on
polylogarithms together with a simple kinematic
map~\cite{Dixon:2021tdw}.  At four points, the MHV form factor obeys an
analogous antipodal self-duality~\cite{Dixon:2022rse}.  We show that the kinematic map underlying this four-point self-duality has
a direct cluster realization.  Restricting the
kinematics to the parity-preserving surface induces a further folding of the
$\widetilde{\mathrm{VI}}$ quiver.  In the resulting rank-four cluster
algebra, the antipodal map is represented by a pair of commuting mutations.
To display the relation between the two lower-point systems at the level of
kinematic boundaries, we find an alternative quiver within the same mutation
class that becomes self-antipodal after folding, mirroring the antipodal
self-duality of the four-point MHV form factor.  Its double- and
triple-collinear boundaries reproduce the $C_2$ algebra of the three-point
form factor and the $A_3$ algebra of the six-point amplitude, respectively.
Folding the $A_3$ boundary to $C_2$ makes the antipodal exchange of the two
$C_2$ boundaries manifest.

Before turning to our main construction, we briefly review periodic momentum twistors
and the cluster algebra conventions used throughout the paper.

\input{figures/periodic_twistors_four_point_manuscript.tex}

\paragraph{Periodic momentum twistors.}
Let $p_i$, $i=1,\ldots,n$, be the massless external momenta and introduce dual
coordinates $x_i$ and spinor-helicity variables
$\{\lambda_i,\widetilde\lambda_i\}$ through
\begin{equation}
 p_i=x_{i+1}-x_i=\lambda_i\widetilde\lambda_i,
 \qquad
 q=\sum_{i=1}^{n}p_i,
 \qquad
 x_{i+n}=x_i+q.
 \label{eq:intro-momentum-twistor}
\end{equation}
Shifted labels are not reduced modulo $n$ but denote image points on the
periodic contour.  We choose $\lambda_{i+n}=\lambda_i$.  The momentum twistors
are
\begin{equation}
 Z_i=(\lambda_i^\alpha,\mu_i^{\dot\alpha})\in\mathbb P^3,
 \qquad
 \mu_i^{\dot\alpha}
 =x_i^{\alpha\dot\alpha}\lambda_{i\alpha}
 =x_{i+1}^{\alpha\dot\alpha}\lambda_{i\alpha},
 \qquad
 Z_{i+n}=\cP Z_i,
 \label{eq:intro-periodic-window}
\end{equation}
where $\cP$ is the projective transformation induced by
$x_i\mapsto x_i+q$.  In momentum twistor space, the dual point $x_i$
corresponds to the line $(Z_{i-1}Z_i)$, and homogeneous four-brackets are
denoted by $\br{ijkl}=\det(Z_i,Z_j,Z_k,Z_l)$.  The four-point periodic contour
and its associated momentum twistors are illustrated in
Fig.~\ref{fig:intro-periodic-twistors}.  For explicit calculations, we use
both the OPE parametrization and a positive $\mathcal X$-coordinate
parametrization; details are given in
Appendix~\ref{app:periodic-parametrization}.

\paragraph{Cluster algebra.}
A seed consists of an ordered list of $\mathcal A$-coordinates and an
extended exchange matrix.  We take $r$ coordinates $A_1,\ldots,A_r$ to be
mutable and $s$ coordinates $f_1,\ldots,f_s$ to be frozen, ordered as
$\{a_1,\ldots,a_{r+s}\}=\{A_1,\ldots,A_r,f_1,\ldots,f_s\}$.  With this
ordering, the extended exchange matrix
$B=(b_{ij})$ has size $(r+s)\times r$.  Its entry $b_{ij}$ records the
oriented connection between node $i$ and mutable node $j$:
$b_{ij}>0$ for $i\to j$ and $b_{ij}<0$ for $j\to i$, while $|b_{ij}|$
gives the arrow multiplicity in the skew-symmetric case.  Mutation at a
mutable node $k$ replaces $a_k$ by
\begin{equation}
 a_k'
 =\frac{1}{a_k}
 \left(
  \prod_{i=1}^{r+s}a_i^{[b_{ik}]_+}
  +
  \prod_{i=1}^{r+s}a_i^{[-b_{ik}]_+}
 \right),
 \label{eq:intro-a-mutation}
\end{equation}
and transforms the exchange matrix according to
\begin{equation}
 b_{ij}'=
 \begin{cases}
  -b_{ij},&i=k\ \text{or}\ j=k,\\
  b_{ij}+[b_{ik}]_+[b_{kj}]_+
  -[-b_{ik}]_+[-b_{kj}]_+,&\text{otherwise},
 \end{cases}
 \label{eq:intro-b-mutation}
\end{equation}
where $[x]_+=\max(x,0)$.  The frozen coordinates
$a_{r+\alpha}=f_\alpha$ do not mutate.  Writing $B_{\mathrm{prin}}$ for the
principal $r\times r$ part of $B$, skew symmetrizability means that there is
a positive diagonal matrix $D$ such that
$B_{\mathrm{prin}}D=-D B_{\mathrm{prin}}^{\mathsf T}$.
The $\mathcal X$-coordinate associated with mutable node $j$ is
\begin{equation}
 x_j=\prod_{i=1}^{r+s}a_i^{b_{ij}}.
 \label{eq:intro-x-coordinate}
\end{equation}
Under mutation at node $k$, these coordinates transform as
\begin{equation}
 x_j'=
 \begin{cases}
  x_k^{-1},&j=k,\\
  x_j\,x_k^{[b_{kj}]_+}(1+x_k)^{-b_{kj}},&j\neq k.
 \end{cases}
 \label{eq:intro-x-mutation}
\end{equation}
To track the rays of the $g$-vector fan, we follow the conventions of
Ref.~\cite{Henke:2021ity} and associate an $r\times r$ $C$-matrix
$C=(c_{ij})$ with each seed and a $g$-vector $\mathbf g_i$ with each mutable
node.  The $j$th column of $C$ is the $c$-vector associated with
node $j$; initially, $C=I_r$ and $\mathbf g_i=\mathbf e_i$.  Under mutation
at node $k$, the $C$-matrix transforms as
\begin{equation}
 c_{ij}'=
 \begin{cases}
  -c_{ij},&j=k,\\
  c_{ij}-[-c_{ik}]_+b_{kj}
  +c_{ik}[-b_{kj}]_+,&j\neq k.
 \end{cases}
 \label{eq:intro-c-mutation}
\end{equation}
Writing $\mathbf b_j^{(0)}$ for the $j$th column of the initial principal
exchange matrix, the $g$-vector at node $k$ mutates as
\begin{equation}
 \mathbf g_k'
  =-\mathbf g_k
  +\displaystyle\sum_{j=1}^{r}[-b_{jk}]_+\mathbf g_j
  +\displaystyle\sum_{j=1}^{r}[c_{jk}]_+\mathbf b_j^{(0)}.
 \label{eq:intro-g-mutation}
\end{equation}
The $g$-vectors in each cluster span a cone of the $g$-vector fan.  Below, we
compare its rays with those of the tropical fan.

\paragraph{Quiver folding.}
Let a quiver admit a symmetry that partitions its nodes into orbits.  We call
the symmetry admissible if no two nodes in a mutable orbit are connected and
all arrows between any two orbits have the same orientation.  Mutations at
the nodes of a mutable orbit then commute.  Identifying the nodes in each
orbit gives a folded
exchange matrix with entries
\begin{equation}
 \overline b_{\alpha\beta}
 =\sum_{i\in\alpha}b_{ij},
 \qquad j\in\beta,
 \label{eq:general-folding}
\end{equation}
where $\alpha$ is any node orbit, $\beta$ is a mutable orbit, and the sum is
independent of the representative $j\in\beta$ by the quiver
symmetry.
Mutation of a folded node is defined by the simultaneous mutations of the
nodes in its orbit.
Two distinct foldings appear in this paper.  \emph{Two-period folding} starts
from a $\Gr(4,2n)$ quiver whose nodes carry periodic momentum twistor
coordinates.  After mutation to a quiver with a $\mathbb Z_2$ symmetry, it
identifies nodes carrying the same function on periodic kinematics.  For
general $n$, we expect one mutable node to be fixed and all others to be
paired, reducing the number of mutable nodes from $(6n-15)$ to $(3n-7)$.
The second operation, \emph{parity folding}, is performed after restricting
the kinematics to the parity-preserving surface and is used to describe the
antipodal map.  It is the cluster algebra folding associated with the
reduction of four-dimensional kinematics to three dimensions
\cite{He:2021eec,Li:2024qct}.

The paper is organized as follows.
Sec.~\ref{sec:g46-c2} constructs the three-point $C_2$ quiver and unfolds
it into a two-period $\Gr(4,6)$ quiver.  Sec.~\ref{sec:g48} obtains the
four-point $\widetilde{\mathrm{VI}}$ cluster algebra from a two-period
$\Gr(4,8)$ quiver
and develops the tropical truncation that gives the
rational and algebraic letters.  Sec.~\ref{sec:antipodal-boundaries} shows
how the antipodal map is represented by mutations and studies the $C_2$ and
$A_3$ algebras arising in the double- and triple-collinear limits.  The final
section summarizes the results and discusses their extension to higher
points.  The appendix gives an explicit OPE parametrization and a positive
$\mathcal X$-coordinate parametrization for four-point periodic momentum
twistors.

\section{A three-point warm-up: unfolding \texorpdfstring{$C_2$}{C2} to a
two-period \texorpdfstring{$\Gr(4,6)$}{Gr(4,6)}}
\label{sec:g46-c2}

At three points, the form factor alphabet exhibits a $C_2$ cluster structure.
Unfolding its quiver and then performing a single mutation produces a
$\Gr(4,6)$ quiver with two periods of frozen nodes.  We now describe this
construction in detail.

On the parity-preserving surface, the six-point amplitude alphabet admits the
$C_2$ parametrization~\cite{Chicherin:2020umh}
\begin{equation}
 \{a_1,a_2,a_3,a_4,a_5,a_6\}
 =\left\{
 \sqrt{\frac{\widehat u}{\widehat v\widehat w}},
 \sqrt{\frac{1-\widehat v}{\widehat v}},
 \sqrt{\frac{\widehat w}{\widehat u\widehat v}},
 \sqrt{\frac{1-\widehat u}{\widehat u}},
 \sqrt{\frac{\widehat v}{\widehat u\widehat w}},
 \sqrt{\frac{1-\widehat w}{\widehat w}}
 \right\},
\label{eq:g46-amplitude-c2}
\end{equation}
where $(\widehat u,\widehat v,\widehat w)$ are the usual six-point
dual-conformal cross-ratios and the parity-preserving surface is defined by
$\widehat\Delta=(1-\widehat u-\widehat v-\widehat w)^2
-4\widehat u\widehat v\widehat w=0$.
The six $a_i$ are the cluster variables of the coefficient-free $C_2$ algebra
and obey the following exchange relations:
\begin{equation}
 a_{m+1}a_{m-1}
 =\begin{cases}
    1+a_m,&m\ \text{odd},\\[2pt]
    1+a_m^2,&m\ \text{even},
  \end{cases}
 \qquad a_{m+6}=a_m.
 \label{eq:g46-c2-recurrence}
\end{equation}
Three-point form factor kinematics is described by the dimensionless ratios
\begin{equation}
 u=\frac{(p_1+p_2)^2}{q^2},\qquad
 v=\frac{(p_2+p_3)^2}{q^2},\qquad
 w=\frac{(p_3+p_1)^2}{q^2},
 \qquad u+v+w=1.
\label{eq:g46-ff-ratios}
\end{equation}
The antipodal map~\cite{Dixon:2021tdw} relates these ratios to the six-point
cross-ratios by
\begin{equation}
 \widehat u=\frac{vw}{(1-v)(1-w)},\qquad
 \widehat v=\frac{uw}{(1-u)(1-w)},\qquad
 \widehat w=\frac{uv}{(1-u)(1-v)}.
 \label{eq:g46-antipodal-kinematic-map}
\end{equation}
The kinematic map identifies the parity-preserving surface
$\widehat\Delta=0$ with the three-point locus $u+v+w=1$.  Consequently,
Eq.~\eqref{eq:g46-amplitude-c2}
gives the following manifest $C_2$ representation of the three-point
form factor alphabet:
\begin{equation}
 \begin{aligned}
 &\{a_1,a_2,a_3,a_4,a_5,a_6\}
 =\left\{
 \frac{1-u}{u},
 \sqrt{\frac{v}{uw}},
 \frac{1-w}{w},
 \sqrt{\frac{u}{vw}},
 \frac{1-v}{v},
 \sqrt{\frac{w}{uv}}
 \right\}\\[5pt]
&=\left\{
 \frac{\br{6234}}{\br{3456}},
 \sqrt{\frac{\br{1234}\br{2356}}{\br{2345}\br{3456}}},
 \frac{\br{5123}}{\br{2345}},
 \sqrt{\frac{\br{3456}\br{1245}}{\br{1234}\br{2345}}},
 \frac{\br{7345}}{\br{1234}},
 \sqrt{\frac{\br{2345}\br{3467}}{\br{3456}\br{1234}}}
 \right\},
 \end{aligned}
\label{eq:g46-ff-c2}
\end{equation}
The second equality follows from the three-point periodic momentum twistor
parametrization.
The signs of the square-root coordinates are chosen consistently with this
parametrization.

\input{figures/g46_c2_seed.tex}

\input{figures/g46_c2_exchange_hexagon.tex}

To construct the corresponding quiver, we take the three consecutive
four-brackets $\br{i\,i{+}1\,i{+}2\,i{+}3}$, $i=1,2,3$, to be frozen; the
choice of $\mathcal A$-coordinates for the mutable nodes is then motivated
by the bracket expressions in Eq.~\eqref{eq:g46-ff-c2}.  The resulting quiver
is shown on the left of Fig.~\ref{fig:g46-c2-seed}.  Its extended exchange
matrix, with rows and columns ordered as displayed, is
\begin{equation}
\begin{pNiceMatrix}[first-row,first-col]
 & \br{5123} & \sqrt{\br{1245}} \\
 \br{5123}        &  0 &  1 \\
 \sqrt{\br{1245}} & -2 &  0 \\ \hline
 \br{1234}        &  1 &  0 \\
 \br{2345}        &  1 & -1 \\
 \br{3456}        & -1 &  0
\end{pNiceMatrix}.
\label{eq:g46-initial-extended-matrix}
\end{equation}
The connection with the form factor alphabet is most transparent in the
associated $\mathcal X$-coordinates.  For the initial quiver, these are
\begin{equation}
 \{x_1,x_2\}
 =\left\{
   \frac{\br{1234}\br{2345}}{\br{3456}\br{1245}},
   \frac{\br{5123}}{\br{2345}}
  \right\}
 =\left\{\frac{vw}{u},\frac{1-w}{w}\right\}.
\label{eq:g46-initial-x-coordinates}
\end{equation}
Mutating the two mutable nodes in the initial quiver gives
\begin{equation}
 \br{5123}
 \longmapsto
 \frac{
   \br{1234}\br{2345}+\br{3456}\br{1245}}
  {\br{5123}}
 =\br{7345},
 \label{eq:g46-upper-mutation}
\end{equation}
and
\begin{equation}
 \sqrt{\br{1245}}
 \longmapsto
 \frac{\br{2345}+\br{5123}}{\sqrt{\br{1245}}}
 =\sqrt{\br{2356}}.
\label{eq:g46-lower-mutation}
\end{equation}
Alternating mutations generate the six mutable $\mathcal A$-coordinates
\begin{equation}
 \left\{
 \br{5123},\sqrt{\br{1245}},
 \br{7345},\sqrt{\br{3467}},
 \br{6234},\sqrt{\br{2356}}
 \right\}.
 \label{eq:g46-complete-coordinate-orbit}
\end{equation}
Each adjacent pair in this sequence, including the last and first
entries, forms one of the six clusters.  These six clusters form the $C_2$
exchange graph shown in Fig.~\ref{fig:g46-c2-exchange-hexagon}, and their
$\mathcal X$-coordinates collectively encode the complete three-point
form factor alphabet.

We next unfold the $C_2$ quiver into the $A_3$ quiver shown in the middle of
Fig.~\ref{fig:g46-c2-seed}.  We obtain it by replacing the mutable node
labelled by $\br{5123}$ with two copies and taking $\br{1245}$, rather than
$\sqrt{\br{1245}}$, as the central mutable coordinate.  Mutating the
circled node labelled by $\br{5123}$ to $\br{7345}$ produces the quiver on
the right, which is an $A_3\simeq\Gr(4,6)$ quiver with two periods of frozen
nodes.
This example motivates the four-point construction below, where we seek a
$\Gr(4,8)$ quiver with the $\mathbb Z_2$ pairing required for folding.

\section{Four-point alphabet from
\texorpdfstring{$\widetilde{\mathrm{VI}}$}{VI-tilde}:
folding, tropical truncation, and limit rays}
\label{sec:g48}

At four points, the periodic construction leads to an infinite cluster
algebra, whereas the form factor alphabet is finite.  We first obtain the
relevant $\widetilde{\mathrm{VI}}$ cluster algebra by folding a two-period
$\Gr(4,8)$ quiver.  A tropical fan selects the relevant mutations, retaining
a finite set of rational coordinates and four physical limit rays, while the
$A_1^{(1)}$ sequences approaching these rays generate the square roots and
algebraic letters.

\subsection{Kinematics and the alphabet}
\label{sec:g48-target}

A convenient cyclic set of dimensionless variables is
\begin{equation}
 u_i=\frac{(p_i+p_{i+1})^2}{q^2},
 \qquad
 v_i=\frac{(p_i+p_{i+1}+p_{i+2})^2}{q^2},
 \qquad i=1,\ldots,4,
 \label{eq:g48-kinematic-ratios}
\end{equation}
where all indices are understood modulo 4.  Momentum conservation and
$p_i^2=0$ impose three independent relations, which may be taken as
$-u_i+u_{i+2}+v_{i-1}+v_i=1$ for $i=1,2,3$, leaving a
five-dimensional kinematic space~\cite{Dixon:2022rse,He:2026ddp}.
We parametrize the four-point periodic momentum twistors by the OPE
variables $\{T_1,S_1,T_2,S_2,\varphi\}$, with $T_j=e^{-\tau_j}$,
$S_j=e^{\sigma_j}$ and
$\varphi=e^{i\phi_2}$~\cite{Basso:2013vsa,Sever:2020jjx,Sever:2021nsq,Sever:2021xby}.
We use this parametrization for collinear limits and the antipodal map, and a
positive $\mathcal X$-coordinate parametrization for the tropical fan.  Both
parametrizations are reviewed in
Appendix~\ref{app:periodic-parametrization}.

Excluding frozen boundary factors, the four-point form factor alphabet contains
\begin{equation}
  88=64_{\mathrm{rat}}+4_{\mathrm{roots}}+4\times5_{\mathrm{alg}}
  \label{eq:g48-target-count}
\end{equation}
letters~\cite{Dixon:2022rse,He:2026ddp}.
$64_{\mathrm{rat}}$ represent the non-frozen rational letters.  Here and
below, ``rational'' means rational in the function field generated by periodic
momentum-twistor, or equivalently Pl\"ucker, coordinates.  Thus
$\operatorname{tr}_5$ and the parity-odd letters involving it are rational in
this sense, although they are not rational functions of the $u_i,v_i$ alone.
The $4_{\mathrm{roots}}$ square roots, by contrast, generate genuine quadratic
extensions of the periodic Pl\"ucker function field and are counted separately
from the $5_{\mathrm{alg}}$ algebraic letters associated with each extension.
The four physical square roots are associated with the following
discriminants, written in terms of the K\"all\'en function
$\lambda(x,y,z)=x^2+y^2+z^2-2xy-2xz-2yz$:
\begin{equation}
\begin{aligned}
 \Delta_{1a}&=\lambda(u_2,u_4,1),
 &\qquad \Delta_{1b}&=\lambda(u_1,u_3,1),\\
 \Delta_{2a}&=\lambda(u_1u_4,u_2u_3,v_1v_3),
 &\qquad \Delta_{2b}&=\lambda(u_1u_2,u_3u_4,v_2v_4).
\end{aligned}
\label{eq:g48-four-discriminants}
\end{equation}

\subsection{From a two-period \texorpdfstring{$\Gr(4,8)$}{Gr(4,8)} quiver to
the \texorpdfstring{$\widetilde{\mathrm{VI}}$}{VI-tilde} cluster algebra}

Motivated by the three-point construction, we start from a $\Gr(4,8)$ quiver
with two periods of frozen nodes, shown in
Fig.~\ref{fig:g48-extended-seed}.  Its arrow structure is inherited from the
standard $\Gr(4,8)$ quiver, whereas its coordinates are evaluated on
four-point periodic momentum twistor kinematics.  Requiring the mutations
used below to remain ordinary four-brackets leads us to replace the coordinate
at node $6$, $\br{1267}$, by $\br{2356}$.  This replacement is compatible
with projective covariance, since the two brackets carry the same twistor
weights on the periodic configuration.

\input{figures/g48_periodicized_extended_seed.tex}

\input{figures/g48_foldable_extended_quiver.tex}

The initial quiver in Fig.~\ref{fig:g48-extended-seed} does not yet exhibit
the symmetry required for folding.  Starting from this quiver, we apply
$\mu_3$, $\mu_6$, $\mu_7$ and $\mu_8$ in that order.  The mutable coordinate
at each mutated node transforms as
\begin{equation}
 \br{1237}\xrightarrow{\mu_3}\br{2456},\quad
 \br{2356}\xrightarrow{\mu_6}\br{1245},\quad
 \br{1345}\xrightarrow{\mu_7}\br{2456},\quad
 \br{1456}\xrightarrow{\mu_8}\br{1236}.
 \label{eq:g48-symmetric-frame-coordinates}
\end{equation}
The resulting quiver, shown in Fig.~\ref{fig:g48-foldable-extended-quiver},
has a $\mathbb Z_2$ symmetry that leaves node $5$ unchanged and exchanges the
mutable pairs $\{1,9\}$, $\{2,8\}$, $\{3,7\}$ and $\{4,6\}$.  The symmetry
also exchanges each frozen node with its counterpart in the other period.  We
order the folded mutable nodes by the orbits
$\{\{5\},\{1,9\},\{2,8\},\{3,7\},\{4,6\}\}$.  Taking the frozen orbits in
the order $\{\br{1234},\br{2345},\br{3456},\br{4567}\}$ and applying the
folding rule in Eq.~\eqref{eq:general-folding} gives
\begin{equation}
 \overline B
 =\begin{pmatrix}
   \overline B_{\mathrm{prin}}\\
   \overline B_{\mathrm{fr}}
  \end{pmatrix},
 \quad
 \overline B_{\mathrm{prin}}
 =\begin{pmatrix}
  0& 1&-1& 1&-1\\
 -2& 0& 1& 0& 1\\
  2&-1& 0&-1& 0\\
 -2& 0& 1& 0& 1\\
  2&-1& 0&-1& 0
 \end{pmatrix},
 \quad
 \overline B_{\mathrm{fr}}
 =\begin{pmatrix}
  0& 1& 0& 0& 0\\
  0& 0& 0& 1&-1\\
  0& 0& 0&-1& 0\\
  0& 0&-1& 1& 0
 \end{pmatrix}.
 \label{eq:g48-periodic-folded-matrix}
\end{equation}
Its principal part obeys
$\overline B_{\mathrm{prin}}D
=-D\overline B_{\mathrm{prin}}^{\mathsf T}$, with
$D=\operatorname{diag}(1,2,2,2,2)$.

\input{figures/g48_folded_quiver.tex}

For the five folded mutable nodes ordered above, mutations at nodes $2$
through $5$ directly replace one four-bracket by another.  At node $1$, the
relevant periodic momentum twistor identity is
\begin{equation}
 \br{1256}\,
 \br{(45)\cap(123)\,2\,(89)\cap(567)\,6}
 =\bigl(\br{1235}\br{2456}-\br{1236}\br{1245}\bigr)^2,
 \label{eq:g48-projected-special-exchange}
\end{equation}
where $(ab)\cap(cde)$ denotes the intersection of the line $(ab)$ with the
plane $(cde)$ and is represented by
$(ab)\cap(cde)=Z_a\br{bcde}-Z_b\br{acde}$.
The perfect square on the right-hand side motivates replacing
$\br{1256}$ by its square root.  We choose the signs so that all exchange
relations take the standard subtraction-free form.  In particular, the sign
of $A_2=-\br{1235}$ converts the difference in
Eq.~\eqref{eq:g48-projected-special-exchange} into the exchange numerator
$A_3A_5+A_2A_4$, while the signs of $F_2$ and $F_4$ make the remaining
exchanges consistent with the same convention.  We therefore use the
following mutable and frozen $\mathcal A$-coordinates:
\begin{equation}
\begin{aligned}
 \{A_1,A_2,A_3,A_4,A_5\}
 &=\{\sqrt{\br{1256}},-\br{1235},\br{1236},
     \br{2456},\br{1245}\},\\
 \{F_1,F_2,F_3,F_4\}
 &=\{\br{1234},-\br{2345},\br{3456},-\br{4567}\}.
\end{aligned}
 \label{eq:g48-lifted-coordinates}
\end{equation}
With these coordinates, all five folded mutations are realized as ordinary
cluster mutations by transposing the principal part of the folded matrix
$\overline B$ and reversing its frozen rows:
\begin{equation}
 B_{\widetilde{\mathrm{VI}}}
 =\begin{pmatrix}
   \overline B_{\mathrm{prin}}^{\mathsf T}\\
   -\overline B_{\mathrm{fr}}
 \end{pmatrix}.
 \label{eq:g48-vi-extended-matrix}
\end{equation}
Together with the $\mathcal A$-coordinates in
Eq.~\eqref{eq:g48-lifted-coordinates}, this extended exchange matrix defines
the five-node quiver shown in Fig.~\ref{fig:g48-folded-quiver}.
With this choice, all five mutations take the standard cluster form.
Its principal part satisfies
$B_{\widetilde{\mathrm{VI}},\mathrm{prin}}D
=-D B_{\widetilde{\mathrm{VI}},\mathrm{prin}}^{\mathsf T}$, with
$D=\operatorname{diag}(2,1,1,1,1)$.

In particular, mutation at node $1$ gives
\begingroup
\small
\begin{equation}
 A_1'
 =\frac{A_3A_5+A_2A_4}{A_1}
 =\frac{\br{1236}\br{1245}-\br{1235}\br{2456}}
        {\sqrt{\br{1256}}}
 =\sqrt{\br{(45)\cap(123)\,2\,(89)\cap(567)\,6}}.
\label{eq:g48-lifted-special-exchange}
\end{equation}
\endgroup
The square root in Eq.~\eqref{eq:g48-lifted-special-exchange}
is taken in the positive $\mathcal X$-coordinate chart defined in
Appendix~\ref{app:periodic-parametrization}.  In this chart, all
coordinates in Eq.~\eqref{eq:g48-lifted-coordinates}, as well as the bracket
under the square root, are positive.  Positivity therefore fixes the branch
in Eq.~\eqref{eq:g48-lifted-special-exchange}, whose square reproduces the
periodic momentum twistor identity in
Eq.~\eqref{eq:g48-projected-special-exchange}.

Up to relabelling, $B_{\widetilde{\mathrm{VI}},\mathrm{prin}}$ belongs to the
$\widetilde{\mathrm{VI}}$ mutation class classified in
Ref.~\cite{Felikson:2010hm}.  The frozen rows of
$B_{\widetilde{\mathrm{VI}}}$ provide the coefficient data for its realization
on four-point form factor kinematics.  The resulting cluster algebra is of
finite mutation type but has infinitely many cluster variables.  The square
root introduced here serves only to recast the folded exchange relations in
standard cluster form and should not be confused with any of the four physical
square roots in the form factor alphabet.
The $\mathcal X$-coordinates associated with the initial seed in
Eq.~\eqref{eq:g48-lifted-coordinates} are
\begin{equation}
 \{x_1,x_2,x_3,x_4,x_5\}
 =\left\{
 \frac{A_2A_4}{A_3A_5},
 \frac{A_3A_5}{A_1^2F_1},
 \frac{A_1^2F_4}{A_2A_4},
 \frac{A_3A_5F_3}{A_1^2F_2F_4},
 \frac{A_1^2F_2}{A_2A_4}
 \right\}.
\label{eq:g48-positive-x-chart}
\end{equation}
We define the positive region by $x_i>0$.  These variables are algebraically
independent and define a positive coordinate chart for the five-dimensional
periodic momentum twistor kinematics.  More details are given in
Appendix~\ref{app:periodic-parametrization}.

\subsection{Tropical truncation and the rational letters}
\label{sec:rational-truncation}

The $\widetilde{\mathrm{VI}}$ cluster algebra has infinitely many cluster
variables, whereas the rational part of the four-point form factor alphabet
is finite.  To select the physically relevant coordinates, we follow
Refs.~\cite{Drummond:2019qjk,Drummond:2019cxm,Henke:2019hve} and tropicalize a
finite set of subtraction-free functions in the positive chart of
Eq.~\eqref{eq:g48-positive-x-chart}.  Their common domains of linearity define
a tropical fan, which provides a selection criterion for mutations.  A
mutation is accepted when the tropical direction associated with its new
$g$-vector is a ray of the fan, and rejected otherwise.  The truncation
therefore retains only clusters whose rays are all contained in the tropical
fan.

For the explicit computation, we associate to each subtraction-free
polynomial $P_F(x)$ its Newton polytope $\operatorname{Newt}(P_F)$, defined as
the convex hull of the exponent vectors of its monomials.  Tropicalization
replaces addition by minimum and multiplication by addition, and its regions
of linearity form the normal fan of $\operatorname{Newt}(P_F)$.  For the finite
input set $S$, their common refinement is therefore the normal fan of
\begin{equation}
 \mathcal N(S)=\sum_{F\in S}\operatorname{Newt}(P_F) .
 \label{eq:rational-newton-sum}
\end{equation}
Here the sum denotes the Minkowski sum of the individual Newton polytopes.
An overall sign or a Laurent-monomial rescaling does not change the normal fan.
The resulting polyhedral computations are performed with
\texttt{polymake}~\cite{Assarf:2017polymake}.

The input used for the rational truncation is the union of the cyclic orbits
generated by the following six minors,
\begin{equation}
 S_{\mathrm{rat}}=
 \{\br{1234},\br{1245},\br{1256},
   \br{1235},\br{1236},\br{1237}\}_{\mathrm{cyc}}.
 \label{eq:rational-input}
\end{equation}
This set consists of the two periodic-bracket families
\[
 \br{i\,i{+}1\,j\,j{+}1}
 \qquad\text{and}\qquad
 \br{i\,j{-}1\,j\,j{+}1}.
\]
After fixing their overall signs, its elements are subtraction-free
polynomials $P_F(x)$ in the positive chart.  Their Newton polytopes define a
complete five-dimensional fan with
\begin{equation}
 f(S_{\mathrm{rat}})=(66,458,1124,1138,408).
 \label{eq:rational-fan-counting}
\end{equation}
To implement the selection rule, we explore mutations in all five mutable
directions from the seed in Eq.~\eqref{eq:g48-vi-extended-matrix}.  The fan
rays are expressed in the exponent basis of the positive coordinates $x_i$,
whereas the $g$-vectors are defined with respect to the initial seed.  The
corresponding map to tropical $x$-space is
\begin{equation}
 r_{\mathcal X}(g)=\mathbb R_{\geq0}\!\left(gD^{-1}\right),
 \qquad D=\operatorname{diag}(2,1,1,1,1)\,.
 \label{eq:rational-g-ray}
\end{equation}
A new cluster is retained only when the ray $r_{\mathcal X}(g)$ of the newly
mutated variable coincides with a ray of the fan of $S_{\mathrm{rat}}$.  The
resulting truncated exchange graph contains 340 clusters.

The 66 rays of the fan split into two classes.  Of these, 64 occur as
$g$-vector rays of cluster variables in the truncated exchange graph.  The
corresponding variables give the $64_{\mathrm{rat}}$ non-frozen rational
letters.  The remaining two rays are limit directions associated with
$\sqrt{\Delta_{1a}}$ and $\sqrt{\Delta_{1b}}$, which we study in the next
subsection.  Table~\ref{tab:rational-cyclic-representatives} gives the 17
cyclic representatives of the rational letters.
The square roots in Table~\ref{tab:rational-cyclic-representatives} are rooted
coordinate representatives and should not be confused with the four physical
square roots in the form factor alphabet. 
The cyclic generator shifts all twistor labels simultaneously,
$Z_i\mapsto Z_{i+1}$, with the labels understood periodically.  We use
$(ab)\cap(cde)$ for the intersection of a line and a plane, and
$(abc)\cap(def)$ for the intersection line of two planes.  Explicitly,
$(ab)\cap(cde)=Z_a\br{bcde}-Z_b\br{acde}$ and
$(abc)\cap(def)=(ab)\br{cdef}+(bc)\br{adef}+(ca)\br{bdef}$, where
$(ab)=Z_a\wedge Z_b$ in the second expression.

\begin{table}[h]
\centering
\small
\renewcommand{\arraystretch}{1.25}
\begin{tabular}{@{}>{\centering\arraybackslash}m{0.14\textwidth}
                    @{\hspace{0.5em}}
                    >{\raggedright\arraybackslash}m{0.83\textwidth}@{}}
\toprule
single four-brackets
& $\br{1235},\ \br{1236},\ \br{1237},\ \br{1245},\
  \br{1246},\ \br{1347},\ \sqrt{\br{1256}},\
   \sqrt{\br{1357}}^{\dagger}$\\
\midrule
composite brackets
& $\br{(12)\cap(567)\,3\,4\,6}$,
  $\br{(23)\cap(456)\,6\,7\,9}$,
  $\br{(45)\cap(123)\,2\,6\,7}$,
  $\br{(45)\cap(123)\,6\,8\,9}$,\newline
  $\br{(56)\cap(123)\,7\,8\,9}$,
  $\br{7\,8\,(234)\cap(456)}$,
  $\sqrt{\br{1\,(23)\cap(456)\,5\,(67)\cap(8\,9\,10)}}$,\newline
  $\sqrt{\br{(45)\cap(123)\,2\,(89)\cap(567)\,6}}$,
  $\sqrt{\br{(45)\cap(678)\,2\,(89)\cap(10\,11\,12)\,6}}^{\dagger}$.\\
\bottomrule
\end{tabular}
\caption{Cyclic representatives of the 64 non-frozen rational letters.  The two
representatives marked by $\dagger$ have cyclic orbit size two; all others
have orbit size four.  Thus the table represents
$2\times2+15\times4=64$ distinct factors.}
\label{tab:rational-cyclic-representatives}
\end{table}

\subsection{Limit rays and algebraic letters}
\label{sec:affine-algebraic-letters}

\input{figures/affine_a11_rooted_origin.tex}
\input{figures/affine_a11_mu21_quiver.tex}

The two limit rays identified above are not represented by cluster
variables.  Instead, they are approached by alternating mutations in an
$A_1^{(1)}$ subquiver, and the resulting recurrence determines a quadratic
discriminant and two associated algebraic letters~\cite{Drummond:2019cxm}.
The generic configuration is shown in
Fig.~\ref{fig:a11-rooted-origin}.  In this diagram, $w_0$ and $z_0$ are the
mutable coordinates connected by the double arrow, while $b$ denotes the
product of the remaining mutable coordinates, and $f_w$ and $f_z$ the corresponding
frozen monomials.  Mutating at $w_0$ or $z_0$ produces $z_1$ or $w_1$,
respectively:
\begin{equation}
 z_1w_0=b+f_wz_0^2,
 \qquad
 w_1z_0=b+f_zw_0^2.
 \label{eq:affine-letter-exchange}
\end{equation}
These two choices define the $z$- and $w$-sequences.  Writing $X_n$ for
either $z_n$ or $w_n$, their coordinates beyond the first step satisfy
\begin{equation}
 X_{n+2}=\mathcal P X_{n+1}-\mathcal F X_n,
 \qquad \mathcal F=f_wf_z,
 \qquad
 \mathcal P=\frac{f_zw_0+z_1}{z_0}
 =\frac{f_wz_0+w_1}{w_0}.
 \label{eq:affine-letter-recurrence}
\end{equation}
Here $\mathcal P$ and $\mathcal F$ remain fixed along the sequence.  Once the
differences between successive $g$-vectors stabilize, both sequences approach
the same limit ray,
\begin{equation}
 \mathbf g_\infty
 =\mathbf g(z_{n+1})-\mathbf g(z_n)
 =\mathbf g(w_{n+1})-\mathbf g(w_n).
 \label{eq:affine-letter-limit-ray}
\end{equation}
The characteristic polynomial of
Eq.~\eqref{eq:affine-letter-recurrence} defines the discriminant
\begin{equation}
 \Delta=\mathcal P^2-4\mathcal F.
 \label{eq:affine-letter-discriminant}
\end{equation}
Among the retained configurations below, discriminants on a given limit ray
have the same square class, while different configurations can produce
different pairs of algebraic letters,
\begin{align}
 \phi_z&=
 \frac{z_0\mathcal P-2f_zw_0+z_0\sqrt\Delta}
      {z_0\mathcal P-2f_zw_0-z_0\sqrt\Delta},
 &
 \phi_w&=
 \frac{w_0\mathcal P-2f_wz_0+w_0\sqrt\Delta}
      {w_0\mathcal P-2f_wz_0-w_0\sqrt\Delta}.
 \label{eq:affine-letter-units}
\end{align}
Changing the sign of $\sqrt\Delta$ inverts both letters and does not change the
multiplicative space that they generate.

We now identify such $A_1^{(1)}$ configurations in the
$\widetilde{\mathrm{VI}}$ cluster algebra.  A configuration at minimal
mutation distance from the initial seed is reached by the mutation sequence
$\{\mu_2,\mu_1\}$ and is shown in
Fig.~\ref{fig:affine-a11-mu21-quiver}.  In this seed, nodes $2$ and $4$
form the $A_1^{(1)}$ pair.
Identifying $w_0$ and $z_0$ with nodes $4$ and $2$, respectively, the signs
fixed in Eq.~\eqref{eq:g48-lifted-coordinates} give
\begin{equation}
 \begin{aligned}
 w_0&=\br{2456},\qquad
 z_0=-\br{1246},\qquad
 b=\br{2468}\br{1236}\br{1245},\\
 f_w&=\frac{\br{2345}\br{4567}}{\br{3456}},\qquad
 f_z=\br{1234}.
 \end{aligned}
\label{eq:affine-letter-example}
\end{equation}
Substituting these data into Eq.~\eqref{eq:affine-letter-discriminant} gives
a discriminant in the square class $\Delta_{2a}$.  The corresponding limit
$g$-vector is obtained from two successive elements of the $w$-sequence:
\begin{equation}
 \mathbf g_\infty=\mathbf g(w_2)-\mathbf g(w_1)
 =(-4,2,0,3,0)-(-2,1,0,2,0)=(-2,1,0,1,0).
\label{eq:affine-letter-example-g-vector}
\end{equation}
This vector specifies the $\Delta_{2a}$ limit ray, which is not retained by
$S_{\mathrm{rat}}$ and enters only after the refinement below.
The two limit rays selected by $S_{\mathrm{rat}}$ account for the
$\Delta_{1a}$ and $\Delta_{1b}$ sectors.  To recover the other two physical
sectors, we refine the tropical input to
\begin{equation}
 S_{\mathrm{full}}
 =S_{\mathrm{rat}}\cup\{\br{1357},\br{2468}\}.
 \label{eq:algebraic-full-input}
\end{equation}
Repeating the Newton polytope calculation with this enlarged input gives a
complete five-dimensional tropical fan with
\begin{equation}
 f(S_{\mathrm{full}})=(76,536,1312,1322,472).
 \label{eq:algebraic-full-fan-counting}
\end{equation}
Adding $\br{2468}$ restores the $\Delta_{2a}$ ray, while adding
$\br{1357}$ restores the $\Delta_{2b}$ ray.  The $A_1^{(1)}$ configurations retained
by the refined fan fall into the four limit ray classes listed in
Table~\ref{tab:intrinsic-physical-rays}.  Thus each physical square-root
sector is associated with a unique limit ray,
although several $A_1^{(1)}$ configurations can lie on the same ray.  Their
directions in the tropical $x$-space are obtained from
Eq.~\eqref{eq:rational-g-ray}.

\begin{table}[h]
\centering
\small
\begin{tabular}{ccc}
\toprule
$\mathbf g_\infty$ & square class & \# $A_1^{(1)}$ configurations\\
\midrule
$(2,0,-1,0,-1)$ & $\Delta_{1a}$ &20\\
$(0,-1,1,-1,1)$ & $\Delta_{1b}$ &20\\
$(-2,1,0,1,0)$ & $\Delta_{2a}$ &12\\
$(2,-1,0,-1,0)$ & $\Delta_{2b}$ &12\\
\bottomrule
\end{tabular}
\caption{The four physical limit rays retained by the refined tropical
fan.  The first column lists their limiting $g$-vectors, and the third gives
the number of retained $A_1^{(1)}$ configurations approaching each ray.
Each configuration consists of the cluster coordinates together with the
quiver.}
\label{tab:intrinsic-physical-rays}
\end{table}

Each retained $A_1^{(1)}$ configuration produces two algebraic letters.
Different configurations approaching the same limit ray may give different
pairs.  For each of the 4 rays in
Table~\ref{tab:intrinsic-physical-rays}, however, the letters generated
collectively have the same multiplicative span as the 5 physical algebraic
letters in the corresponding square-root sector.  The 4 rays therefore
account for all 20 algebraic letters of the form factor alphabet.

The $f$-vector in Eq.~\eqref{eq:algebraic-full-fan-counting} shows that the
refined fan has 76 rays, 4 of which are physical limit rays.  The truncated
exchange graph realizes the remaining 72 as cluster variables: the 64
physical rational letters and 8 additional coordinates.  Up to overall signs,
the latter are generated by 2 cyclic
representatives,
\begin{equation}
 \mathcal E_8=
 \left\{
 \br{(89)\cap(567)\,2\,4\,6},
 \br{(12)\cap(345)\,4\,(56)\cap(789)\,7}
 \right\}_{\mathrm{cyc}},
 \label{eq:algebraic-extra-rational-orbits}
\end{equation}
each with cyclic orbit size four.  Thus the refined truncation captures all
four physical algebraic sectors while introducing two additional cyclic
orbits of rational coordinates.  In Sec.~\ref{sec:antipodal-boundaries}, we
show that antipodal duality distinguishes these coordinates from the physical
rational alphabet.

\section{Antipodal duality and kinematic boundaries}
\label{sec:antipodal-boundaries}

We now use the $\widetilde{\mathrm{VI}}$ cluster algebra to study antipodal
duality and collinear boundaries.  On the parity-preserving surface, a further
folding gives a rank-four cluster algebra in which the antipodal map is represented
by a pair of commuting mutations.  Closure under this map retains the
physical rational alphabet while excluding the 8 additional coordinates.
The double- and triple-collinear limits reproduce the $C_2$ algebra of the
three-point form factor and the $A_3$ algebra of the six-point amplitude,
respectively.  To make their relation manifest, we use an alternative quiver
in the same mutation class.  On the parity-preserving surface, its $A_3$
boundary folds to $C_2$, and the resulting quiver is self-antipodal.  The
antipodal action then exchanges this parity-folded $C_2$ boundary with the
$C_2$ boundary of the three-point form factor.

\subsection{The antipodal map as a cluster symmetry}
\label{sec:antipodal-cluster}

\input{figures/antipodal_folded_quiver.tex}

To compare the antipodal map with cluster mutations, we use the five
$\mathcal X$-coordinates $\{x_1,\ldots,x_5\}$ of
Eq.~\eqref{eq:g48-positive-x-chart}.  In the OPE parametrization, parity
preservation sets $\varphi=1$, equivalently
$\operatorname{tr}_5=0$~\cite{Dixon:2022rse}.  The explicit parametrization
then gives
\begin{equation}
 \frac{x_2}{x_4}=\varphi^2,
\end{equation}
and hence $x_2=x_4$ on this surface.  The principal exchange matrix is
compatible with this identification: it is invariant under the interchange
of nodes $2$ and $4$.  We may therefore fold the five-node
$\widetilde{\mathrm{VI}}$ quiver according to the orbits $\{1\}$,
$\{2,4\}$, $\{3\}$ and $\{5\}$.  Writing the folded $\mathcal X$-coordinates
in this orbit order gives
\begin{equation}
 \{\bar x_1,\bar x_2,\bar x_3,\bar x_4\}
 =\{x_1,x_2(=x_4),x_3,x_5\}.
\end{equation}
The resulting four-node quiver and its principal exchange matrix are shown in
Fig.~\ref{fig:antipodal-folded-quiver}.

On the parity-preserving surface, the antipodal map acts on the OPE
variables as~\cite{Dixon:2022rse}
\begin{equation}
 g:\quad
 (T_1,S_1,T_2,S_2)
 \longmapsto
 \left(
  \sqrt{\frac{T_2}{S_2}},
  \sqrt{\frac{1}{T_2S_2}},
  \frac{T_1}{S_1},
  \frac{1}{T_1S_1}
 \right).
 \label{eq:antipodal-ope-map}
\end{equation}
Its induced action on the folded $\mathcal X$-coordinates is
\begin{equation}
 g(\bar x)=
 \left\{
  \frac{\bar x_2\bar x_3\bar x_4}
       {(1+\bar x_3)(1+\bar x_4)},
  \bar x_1(1+\bar x_3)(1+\bar x_4),
  \frac1{\bar x_4},\frac1{\bar x_3}
 \right\}.
 \label{eq:antipodal-cluster-x-action}
\end{equation}
The same functions are obtained by mutating nodes $4$ and $3$ successively:
\begin{equation}
\begin{aligned}
 \{\bar x_1,\bar x_2,\bar x_3,\bar x_4\}
 &\xrightarrow{\ \mu_4\ }
 \left\{
  \bar x_1(1+\bar x_4),
  \frac{\bar x_2\bar x_4}{1+\bar x_4},
  \bar x_3,
  \frac1{\bar x_4}
 \right\}\\
 &\xrightarrow{\ \mu_3\ }
 \left\{
  \bar x_1(1+\bar x_3)(1+\bar x_4),
  \frac{\bar x_2\bar x_3\bar x_4}
       {(1+\bar x_3)(1+\bar x_4)},
  \frac1{\bar x_3},
  \frac1{\bar x_4}
 \right\}.
\end{aligned}
 \label{eq:antipodal-cluster-word}
\end{equation}
Up to relabelling of the nodes, the final set is precisely $g(\bar x)$.
Thus, on the parity-preserving surface, the antipodal map is represented
by the cluster transformation $\mu_3\mu_4$ of the folded
$\widetilde{\mathrm{VI}}$ quiver.

The antipodal map also provides a criterion for removing the 8 additional
rational coordinates introduced by $S_{\mathrm{full}}$.  Let
$\mathcal R_{68}$ denote the extended rational factor set, consisting of the
64 physical mutable factors and 4 frozen boundary factors.  On the
parity-preserving surface, exact factorization of their antipodal images gives
\begin{equation}
 g\!\left(\langle\mathcal R_{68}\rangle_{\mathrm{mult}}\right)
 =\langle\mathcal R_{68}\rangle_{\mathrm{mult}},
 \qquad
 g(E)\notin\langle\mathcal R_{68}\cup\mathcal E_8\rangle_{\mathrm{mult}}
 \quad(E\in\mathcal E_8),
 \label{eq:antipodal-rational-selection}
\end{equation}
where $\langle\cdot\rangle_{\mathrm{mult}}$ denotes the multiplicative space
generated by the indicated functions, allowing rational powers.  Thus the
extended set $\mathcal R_{68}$ is closed under the antipodal map, whereas the
image of every additional coordinate contains a new irreducible factor
outside $\mathcal R_{68}\cup\mathcal E_8$.  Applied to the finite tropical
set, requiring antipodal closure therefore retains the 64 physical rational
letters and excludes all eight additional coordinates; the 4 frozen boundary
factors remain as coefficients.  On the parity-preserving surface, each
algebraic sector reduces to a
four-dimensional multiplicative space.  The antipodal map pairs these spaces
according to $\Delta_{1a}\leftrightarrow\Delta_{1b}$ and
$\Delta_{2a}\leftrightarrow\Delta_{2b}$.

\subsection{\texorpdfstring{Collinear boundaries: $A_3$ and $C_2$}
{Collinear boundaries: A3 and C2}}
\label{sec:kinematic-boundaries}

\input{figures/triple_collinear_mutated_quiver.tex}

Having identified the antipodal map as a cluster transformation, we now
consider two collinear boundaries of the four-point kinematics.  The
double-collinear limit directly reproduces the $C_2$ cluster algebra of the
three-point form factor.  The triple-collinear limit yields the $A_3$ cluster
algebra of six-point kinematics.  On the parity-preserving surface,
identifying the parity-related nodes folds this $A_3$ boundary to a $C_2$
boundary quiver.  The antipodal map exchanges this $C_2$ boundary with the
double-collinear one.

To display the relation between the two collinear boundaries in a single
quiver, we successively mutate nodes $3$, $1$, $2$ and $4$ of the initial
$\widetilde{\mathrm{VI}}$ quiver, where the last two mutations commute.  The
resulting quiver is shown in the left panel of
Fig.~\ref{fig:triple-collinear-mutated-quiver}.  We denote the
$\mathcal X$-coordinates on this quiver again by $x_1,\ldots,x_5$.  Its
nodes $\{1,3\}$ form the double-collinear $C_2$ boundary, while nodes
$\{2,4,5\}$ form the
triple-collinear $A_3$ boundary.  On the parity-preserving surface,
identifying nodes $2$ and $4$ folds the latter to a second $C_2$ boundary.
In this folded quiver, the antipodal map acts by exchanging
$1\leftrightarrow\{2,4\}$ and $3\leftrightarrow5$, and therefore directly
exchanges the two $C_2$ boundary subquivers.

\paragraph*{Double-collinear boundary.}
\phantomsection
\label{sec:double-collinear-boundary}

Consider the strict double-collinear limit $T_2\to0$.  At leading power, the
four-point form factor reduces to the three-point form factor, whose
kinematics depend only on $\{T_1,S_1\}$.  For the quiver in the left panel of
Fig.~\ref{fig:triple-collinear-mutated-quiver}, the five
$\mathcal X$-coordinates behave as
\begin{equation}
 \{x_1,x_2,x_3,x_4,x_5\}
 \sim
 \left\{
  \frac{1}{S_1^2+T_1^2},
  \frac{S_2(S_1^2+T_1^2)}{\varphi S_1^2}\,T_2,
  \frac{S_1^2}{T_1^2},
  \frac{\varphi S_2(S_1^2+T_1^2)}{S_1^2}\,T_2,
  T_2^{-2}
 \right\}.
\end{equation}
Thus $x_2$ and $x_4$ vanish, while $x_5$ diverges.  The coordinates on nodes
$1$ and $3$ remain finite and nonzero and depend only on the reduced
three-point kinematics.  These two nodes therefore form the $C_2$ boundary
subquiver highlighted in orange in
Fig.~\ref{fig:triple-collinear-mutated-quiver}.

Let $\{u,v,w\}$, with $u+v+w=1$, be the three cyclic ratios of the
three-point form factor.  The corresponding six $\mathcal X$-coordinates may
be written as
\begin{equation}
 \{\mathfrak a,\mathfrak b,\mathfrak c,\mathfrak d,\mathfrak e,\mathfrak f\}
 =\left\{
 \frac{u}{vw},\frac{v}{wu},\frac{w}{uv},
 \frac{1-u}{u},\frac{1-v}{v},\frac{1-w}{w}
 \right\}.
 \label{eq:three-point-six-letters}
\end{equation}
The surviving coordinates are
\begin{equation}
 \lim_{T_2\to0}\left\{x_1,x_3\right\}
 =\{\mathfrak e^{-1},\mathfrak a\}.
\end{equation}
This is the upper-right cluster in
Fig.~\ref{fig:g46-c2-exchange-hexagon}, with the two coordinates displayed
there in the opposite order.
Alternating mutations of the two surviving nodes generate the complete
$C_2$ exchange hexagon,
\begin{equation}
 \begin{split}
 \{\mathfrak e^{-1},\mathfrak a\}
 &\xrightarrow{\mu_1}\{\mathfrak e,\mathfrak c^{-1}\}
 \xrightarrow{\mu_3}\{\mathfrak d^{-1},\mathfrak c\}
 \xrightarrow{\mu_1}\{\mathfrak d,\mathfrak b^{-1}\}\\
 &\xrightarrow{\mu_3}\{\mathfrak f^{-1},\mathfrak b\}
 \xrightarrow{\mu_1}\{\mathfrak f,\mathfrak a^{-1}\}
 \xrightarrow{\mu_3}\{\mathfrak e^{-1},\mathfrak a\}.
 \end{split}
 \label{eq:double-collinear-C2-orbit}
\end{equation}
The double-collinear boundary therefore reproduces the complete $C_2$
exchange graph of the three-point form factor.

\paragraph*{Triple-collinear boundary.}
\phantomsection
\label{sec:triple-collinear-boundary}

We next consider the strict triple-collinear limit $T_1\to0$.  The reduced
six-point kinematics depend only on $\{T_2,S_2,\varphi\}$, while $S_1$ drops
out.  For the quiver in the left panel of
Fig.~\ref{fig:triple-collinear-mutated-quiver}, the five
$\mathcal X$-coordinates behave as
\begin{equation}
 \{x_1,x_2,x_3,x_4,x_5\}\sim
 \left\{
 \frac{1+T_2^2}{S_1^2},\,
 \frac{S_2T_2}{\varphi(1+T_2^2)},\,
 \frac{S_1^2}{T_1^2},\,
 \frac{\varphi S_2T_2}{1+T_2^2},\,
 T_2^{-2}
 \right\},
 \label{eq:triple-collinear-five-x-leading}
\end{equation}
Thus $x_3$ diverges, while $x_1$, although finite, retains dependence on
$S_1$ and is absent from the reduced six-point kinematics.  The coordinates
on nodes $2$, $4$ and $5$ are finite and independent of $S_1$.  These three
nodes form the $A_3$ boundary subquiver highlighted in blue in
Fig.~\ref{fig:triple-collinear-mutated-quiver}.  Its
identification with the standard cluster algebra of six-point kinematics is
made explicit by the standard OPE parametrization of six-point kinematics
of Ref.~\cite{Basso:2013vsa}:
\begin{equation}
 \widehat u=
 \frac{1}{1+(T_2+S_2\varphi)(T_2+S_2/\varphi)},\qquad
 \widehat v=\frac{\widehat u S_2^2}{1+T_2^2},\qquad
 \widehat w=\frac{T_2^2}{1+T_2^2}.
 \label{eq:triple-collinear-hexagon-uvw}
\end{equation}
The corresponding discriminant is
\begin{equation}
 \Delta_6=
 \left[
 \frac{\widehat u S_2T_2(\varphi-\varphi^{-1})}{1+T_2^2}
 \right]^2.
\end{equation}
The relation to the boundary quiver is particularly transparent in the
$\mathcal X$-coordinates.  With the sign of the square root fixed by the OPE
parametrization above, the three coordinates on its initial $A_3$ cluster are
\begin{equation}
 \{x_2,x_4,x_5\}=
 \left\{
 \frac{2\widehat v\widehat w}
      {1-\widehat u-\widehat v-\widehat w+\sqrt{\Delta_6}},\,
 \frac{2\widehat v\widehat w}
      {1-\widehat u-\widehat v-\widehat w-\sqrt{\Delta_6}},\,
 \frac{1-\widehat w}{\widehat w}
 \right\}.
 \label{eq:triple-collinear-initial-cluster-uvw}
\end{equation}
Enumerating the complete exchange graph gives the expected $14$ clusters.
We verify that the multiplicative space generated by their
$\mathcal X$-coordinates is identical to that generated by the standard
9-letter hexagon alphabet~\cite{Dixon:2011pw}.

Parity acts as $\varphi\mapsto\varphi^{-1}$~\cite{Basso:2013vsa}.  Under
this transformation, nodes $2$ and $4$ are exchanged.
On the parity-preserving surface $\varphi=1$, identifying these two nodes
folds the $A_3$ boundary quiver in the left panel of
Fig.~\ref{fig:triple-collinear-mutated-quiver} to the $C_2$ quiver in the
middle panel.  Its two $\mathcal X$-coordinates are
\begin{equation}
 \{x_{\{2,4\}},x_5\}
 =\left\{
  \frac{S_2T_2}{1+T_2^2},
  T_2^{-2}
 \right\}
 =\left\{
  \sqrt{\frac{\widehat v\widehat w}{\widehat u}},
  \frac{1-\widehat w}{\widehat w}
 \right\}.
\end{equation}
The antipodal map exchanges the two collinear boundaries.  Indeed,
the first and third components of Eq.~\eqref{eq:antipodal-ope-map} show that
$T_1\to0$ and $T_2\to0$ are interchanged.  Substituting the same map into
the folded boundary coordinates gives
\begin{equation}
 g:\quad
 \{x_{\{2,4\}},x_5\}
 \longmapsto
 \{\mathfrak e^{-1},\mathfrak a\}.
 \label{eq:antipodal-folded-boundary-image}
\end{equation}
Eq.~\eqref{eq:antipodal-folded-boundary-image} is the restriction of the
antipodal map to the triple-collinear $C_2$ boundary.  More generally, on the
full folded quiver in the middle panel of
Fig.~\ref{fig:triple-collinear-mutated-quiver}, the antipodal map acts by
\begin{equation}
 g:\quad 1\leftrightarrow\{2,4\},\qquad 3\leftrightarrow5.
\end{equation}
Thus the folded four-point quiver is self-antipodal: it is invariant under the
antipodal map up to the displayed permutation of nodes, mirroring the
antipodal self-duality of the four-point form factor.  On its collinear boundaries, the
same cluster transformation maps the parity-folded $C_2$ algebra of the
six-point amplitude to the $C_2$ algebra of the three-point form factor,
providing a cluster-algebraic realization of the kinematic map underlying
their antipodal duality.

\section{Summary and outlook}
\label{sec:discussion}

We have developed a cluster algebra description of stress-tensor form factor
alphabets based on periodic momentum twistors.  At three points, the $C_2$
quiver unfolds to a two-period $\Gr(4,6)$ quiver.  Guided by this example,
folding a two-period $\Gr(4,8)$ quiver at four points leads
to the rank-five $\widetilde{\mathrm{VI}}$ cluster algebra.  Although it is of
finite mutation type, the corresponding cluster algebra contains infinitely
many cluster variables.  Tropical truncation selects a finite set containing
all the physical rational coordinates and four limit rays.  The $A_1^{(1)}$
mutation sequences approaching these rays generate the square roots and
algebraic letters.  Within this finite tropical candidate set, antipodal
closure excludes the eight additional rational coordinates introduced when
all four limit rays are retained.

Antipodal duality is also encoded directly in this cluster structure.  On the
parity-preserving surface, a further folding gives a rank-four cluster algebra
in which the antipodal map is represented by two commuting mutations followed
by a relabelling.  The double- and triple-collinear boundaries give the $C_2$
algebra of the three-point form factor and the $A_3$ algebra of the six-point
amplitude, respectively.  After parity folding the latter to $C_2$, the same
cluster automorphism exchanges the two boundary algebras.  This provides a
cluster-algebraic realization of the kinematic maps underlying the antipodal
self-duality of the four-point MHV form factor and the antipodal duality between
its two collinear limits.

Beyond identifying the letters, our construction raises a direct bootstrap
question: which letters can appear consecutively in a symbol?  For the
three-point $C_2$ alphabet, cluster adjacency can be identified with the
extended Steinmann relations~\cite{Chicherin:2020umh}, and related constraints
have played a central role in bootstrap calculations through eight
loops~\cite{Dixon:2020bbt,Dixon:2022xqh}.  At four points, extended Steinmann
relations already help determine the NMHV form factor at three
loops~\cite{He:2026ddp,HeLiuYang:toappear}, but no cluster adjacency rule has
been formulated for the full alphabet of $88$ letters.  The retained clusters
and tropical fan provide concrete candidate compatibility data.  Comparing
these data with adjacent letter pairs in the known symbols can test whether
cluster adjacency imposes constraints beyond extended Steinmann and clarify
how the algebraic letters enter.

Final entries provide a complementary constraint.  A simple
all-multiplicity last-entry pattern for two-loop MHV form factors was found
from the periodic Wilson loop~\cite{Li:2024qtn}, and its higher-loop extension
is under study~\cite{DixonLi:toappear}.  No comparable condition is known for
the four-point NMHV form factor.  An analogue of the amplitude $\bar Q$
equation~\cite{He:2019jee,He:2020vob,Li:2021bwg} would require extending the
dual-supersymmetry anomaly to the periodic super Wilson line.  Together with
adjacency, such a condition could push the current four-point NMHV bootstrap
beyond three loops, with four loops as the first target
\cite{He:2026ddp,HeLiuYang:toappear}.

The periodic geometry is not restricted to weak coupling.  At strong
coupling, form factors are described by minimal surfaces ending on an
infinite periodic null contour and by associated $Y$-systems
\cite{Maldacena:2010kp,Gao:2013dza}.  At finite coupling, the collinear
dynamics of the periodic Wilson line is organized by the form factor OPE
\cite{Sever:2020jjx,Sever:2021nsq,Sever:2021xby}.  It remains to understand
whether the cluster coordinates found here have a direct role in either
description.

Five points provide the next concrete test of the proposed folding
construction based on a two-period $\Gr(4,10)$ quiver.  The
two-loop five-point MHV form factor has a 135-letter alphabet involving ten
square roots: five associated with one-loop triangle roots and five with XC
roots.  Of its letters, 30 involve the former roots and 20 involve the latter,
while the remaining 85 are rational~\cite{Li:2024qtn}.  A complete five-point
construction should account for these algebraic sectors and explain how the
finite physical alphabet is selected from the cluster algebra.  The eight-letter
ambiguity at four points illustrates the issue: folding determines the ambient
cluster structure, but an additional physical criterion is needed to select a
finite alphabet.  Antipodal closure supplies such a criterion at four points;
no analogous antipodal selection rule is presently known at five points.  We do not
attempt a five-point truncation here; instead, we test the proposed construction
through its multi-collinear boundaries.

These boundaries can be compared directly with the factorization limits of
the two-loop form factor~\cite{Li:2024qtn}.  We use $T_i=e^{-\tau_i}$,
$S_i=e^{\sigma_i}$ and $\varphi_i=e^{i\phi_i}$ for the OPE variables.  As
summarized in Table~\ref{tab:g410-collinear-boundaries}, the $T_3\to0$ limit of the folded
$\Gr(4,10)$ quiver contains the four-point
$\widetilde{\mathrm{VI}}$ algebra, consistently with the reduction of the
five-point form factor to the four-point one.  The triple-collinear limit
$T_2\to0$ gives $C_2\times A_3$, corresponding to the three-point form factor
and the six-point amplitude, while the quadruple-collinear limit $T_1\to0$
gives $E_6$, corresponding to the seven-point amplitude.  These
three limits provide nontrivial checks of the proposed five-point quiver.

\begin{table}[h]
 \centering
 \small
 \renewcommand{\arraystretch}{1.25}
 \begin{tabular}{cccc}
  \toprule
  OPE limit & Physical boundary & Boundary subalgebra & Surviving OPE variables \\
  \midrule
  $T_3\to0$ & double-collinear
   & $\widetilde{\mathrm{VI}}$
   & $\{T_1,S_1,T_2,S_2,\varphi_2\}$ \\[5pt]
  $T_2\to0$ & triple-collinear
   & $C_2\times A_3$
   & $\begin{matrix}
       C_2:\ \{T_1,S_1\},\\[-2pt]
       A_3:\ \{T_3,S_3,\varphi_3\}
      \end{matrix}$ \\[12pt]
  $T_1\to0$ & quadruple-collinear
   & $E_6$
   & $\{T_2,S_2,\varphi_2,T_3,S_3,\varphi_3\}$ \\
  \bottomrule
 \end{tabular}
 \caption{Collinear boundary subalgebras obtained from the folded
  $\Gr(4,10)$ quiver.}
 \label{tab:g410-collinear-boundaries}
\end{table}

The higher-point extension of antipodal duality remains open.  At four points,
the cluster symmetry emerges only after restricting to the parity-preserving
surface and performing a further folding of the
$\widetilde{\mathrm{VI}}$ quiver.  The two-period folding by itself does not
provide an analogous antipodal action at general multiplicity.  In particular,
although the folded $\Gr(4,10)$ quiver has the expected multi-collinear
boundary subalgebras, we have not identified an antipodal symmetry of the full
five-point quiver.  Whether such a symmetry exists at higher points, and
whether it distinguishes even from odd multiplicities, remains an open
question.

More generally, it remains to determine for which values of $n$ folding the
periodic $\Gr(4,2n)$ quiver produces a consistent cluster algebra.
When the resulting algebra is infinite, one must also explain how its finite
physical alphabet is selected.  The annular-network construction in
Appendix~\ref{app:periodic-parametrization} gives an explicit positive
$\mathcal X$-coordinate parametrization of four-point periodic momentum
twistor kinematics.  At higher points, the analogous coordinates must be
shown to satisfy the folded exchange relations, including the frozen
coefficients.  Even then, the cluster algebra
provides only candidate symbol letters; which of them actually occur depends
on the observable and the loop order.

\acknowledgments

We thank Qinglin Yang for collaboration on related work, and Lance Dixon,
Xuhang Jiang, Marcus Spradlin, and Anastasia Volovich for inspiring
discussions.
We are also grateful to Lance Dixon for comments on the draft.
OpenAI Codex was used to assist with literature searches, exploratory
calculations and language editing.  The authors reviewed and revised all
AI-assisted material and take full responsibility for the content.  The work
of S.H. is supported by the National Natural Science Foundation of China under
Grant Nos.~12225510 and 12247103, and by the New Cornerstone Science
Foundation.

\appendix

\section{Parametrizations of four-point periodic momentum twistors}
\label{app:periodic-parametrization}

Four-point periodic kinematics is five-dimensional.  We give the two
parametrizations used in the main text: the OPE parametrization for the
collinear limits and the antipodal map, and a positive $\mathcal X$-coordinate
parametrization for the tropical calculation.

\paragraph{OPE parametrization.}
We use the variables $\{T_1,S_1,T_2,S_2,\varphi\}$, where
$T_i=e^{-\tau_i}$, $S_i=e^{\sigma_i}$ and $\varphi=e^{i\phi_2}$
\cite{Basso:2013vsa,Sever:2020jjx,Sever:2021nsq,Sever:2021xby}.  Following
Ref.~\cite{Li:2024qtn}, we use the following matrices:
\begin{equation}
 M_1=\operatorname{diag}(T_1,T_1^{-1},S_1,S_1^{-1})
 \label{eq:app-ope-m1}
\end{equation}
and
\begin{equation}
 M_2=
 \begin{pmatrix}
 S_2&0&0&0\\
 -S_2(T_2^2-1)&S_2T_2^2&0&S_2T_2(\varphi S_2-T_2)\\
 \varphi T_2-S_2&0&\varphi T_2&0\\
 0&0&0&\varphi S_2^2T_2
 \end{pmatrix},
 \label{eq:app-ope-m2}
\end{equation}
where an overall factor in $M_2$ has been omitted because the twistors are
homogeneous.
The four twistors in one period are
\begin{equation}
 \begin{aligned}
 Z_1&=M_1(0,1,0,1)^{\mathsf T},&
 Z_2&=M_1M_2(1,3,-1,1)^{\mathsf T},\\
 Z_3&=M_1M_2(1,1,-2,0)^{\mathsf T},&
 Z_4&=(0,0,1,0)^{\mathsf T}.
 \end{aligned}
 \label{eq:app-ope-twistors}
\end{equation}
Their periodic images are defined by $Z_{i+4}=\mathcal P_{\mathrm{OPE}}Z_i$,
with
\begin{equation}
 \mathcal P_{\mathrm{OPE}}=
 \begin{pmatrix}
 2&1&0&0\\
 -1&0&0&0\\
 0&0&2&1\\
 0&0&-1&0
 \end{pmatrix}.
 \label{eq:app-ope-monodromy}
\end{equation}
The parity-preserving surface is $\varphi=1$.  The double-collinear and
triple-collinear limits correspond to $T_2\to0$ and $T_1\to0$, respectively.

\paragraph{Positive $\mathcal X$-coordinate parametrization.}
We next construct the positive chart of
Eq.~\eqref{eq:g48-positive-x-chart} from a bipartite network.  The two
networks used in the construction are shown in
Fig.~\ref{fig:periodic-bipartite-networks}.

\input{figures/periodic_plabic_graphs}

The disk network in the left panel is dual to the symmetric $\Gr(4,8)$
quiver of Fig.~\ref{fig:g48-foldable-extended-quiver}.  Its central face,
labelled by $x_1$, is fixed by the folding, whereas the remaining faces are
paired as indicated by their labels and colors.  Excising the fixed point
from the central face and taking the quotient gives the annular network in
the right panel.  The quotient network has five independent face variables
$\{x_1,\ldots,x_5\}$.  The dashed curve is a cut from the puncture to
the outer boundary; a path crossing it acquires a factor of $\lambda$, so
the power of $\lambda$ records the winding number in the annulus.

The arrows in the right panel specify a perfect orientation with boundary
sources $1,2$ and sinks $3,4$, and the edges carry the positive weights
$\alpha_1,\ldots,\alpha_{18}$ shown in the figure.  We use the standard
boundary measurement for directed networks in a disk and an annulus
\cite{Postnikov:2006,Gekhtman:2009}.  The edge weights are defined up to
rescalings at internal vertices.  For each face $f$, define
\begin{equation}
 x_f=\prod_{e\subset\partial f}\alpha_e^{\epsilon_{f,e}},
 \qquad
 \epsilon_{f,e}=
 \begin{cases}
  +1,& e\text{ points counterclockwise around }f,\\
  -1,& e\text{ points clockwise around }f.
 \end{cases}
 \label{eq:app-face-variable}
\end{equation}
For the network in the right panel,
\begin{equation}
 \begin{aligned}
  x_1&=\frac{\alpha_6\alpha_7\alpha_8}{\alpha_9},
  &x_2&=\frac{\alpha_9\alpha_{10}\alpha_{11}
                    \alpha_{14}\alpha_{16}}{\alpha_{12}},\\[2pt]
  x_3&=\frac{\alpha_2}
              {\alpha_3\alpha_8\alpha_{10}},
  &x_4&=\frac{\alpha_1}
              {\alpha_2\alpha_5\alpha_7},
  &x_5&=\frac{\alpha_5}
              {\alpha_6\alpha_{14}\alpha_{15}}.
 \end{aligned}
 \label{eq:app-annular-face-variables}
\end{equation}
We choose the following representative of these face weights:
\begin{equation}
 \alpha_1=x_1x_3x_4x_5,
 \qquad \alpha_2=x_1x_3,
 \qquad \alpha_5=x_5,
 \qquad \alpha_8=x_1,
 \qquad \alpha_{11}=x_2.
 \label{eq:app-edge-weight-section}
\end{equation}
All remaining edge weights are set to $1$.
For a directed path $p$ from source $i$ to boundary vertex $j$, define
\begin{equation}
 \operatorname{wt}(p)
 =\lambda^{\operatorname{wind}(p)}
  \prod_{e\in p}\alpha_e,
 \qquad
 M_{ij}(\lambda)
 =\sum_{p:i\to j}\operatorname{wt}(p),
 \label{eq:app-directed-path-weight}
\end{equation}
where $\operatorname{wind}(p)$ is its signed number of crossings of the
annular cut, and the sum runs over all directed paths from $i$ to $j$.
In the representative of Eq.~\eqref{eq:app-edge-weight-section}, the resulting
boundary measurement matrix is
\begin{equation}
 M(\lambda)=
 \begin{pmatrix}
 1&0&x_2(\lambda+x_1+x_1x_5)&
       \lambda+\lambda x_2+x_1x_2\\
 0&1&\lambda+x_1+x_1x_3+x_1x_5+x_1x_3x_5+x_1x_3x_4x_5&
       \lambda+x_1+x_1x_3
 \end{pmatrix}.
 \label{eq:app-periodic-boundary-measurement}
\end{equation}
Every entry is a subtraction-free polynomial in
$\{x_1,\ldots,x_5,\lambda\}$.  The sign associated with the cyclic ordering
of the two sources is included by setting
\begin{equation}
 \widehat M(\lambda)=M(\lambda)
 \operatorname{diag}(1,1,1,-1).
 \label{eq:app-periodic-boundary-sign}
\end{equation}
This fixed column sign does not alter the positive path sums.

The matrix $\widehat M$ has two rows, while a momentum twistor has four
components.  We use its first jet at $\lambda=1$ to lift the annular boundary
measurement to a four-dimensional periodic momentum twistor configuration,
retaining both the value of each column and its first logarithmic derivative.
A shift
by $n$ periods multiplies the boundary measurement by $\lambda^n$; hence,
writing $\widehat M_j$ for the $j$th column, we define
\begin{equation}
 Z_{4n+j}=
 \begin{pmatrix}
  \widehat M_j(1)\\[2pt]
  \left.(\lambda\partial_\lambda)
  \bigl(\lambda^n\widehat M_j(\lambda)\bigr)\right|_{\lambda=1}
 \end{pmatrix},
 \qquad j=1,\ldots,4,\qquad n\in\mathbb Z.
 \label{eq:app-periodic-first-jet}
\end{equation}
For one period, this construction gives
\begin{align}
 Z_1&=(1,0,0,0)^{\mathsf T},
 &Z_2&=(0,1,0,0)^{\mathsf T},
 \nonumber\\
 Z_3&=\begin{pmatrix}
 x_2(1+x_1+x_1x_5)\\
 1+x_1+x_1x_3+x_1x_5+x_1x_3x_5+x_1x_3x_4x_5\\
 x_2\\1
 \end{pmatrix},
 &Z_4&=-\begin{pmatrix}
 1+x_2+x_1x_2\\
 1+x_1+x_1x_3\\
 1+x_2\\1
 \end{pmatrix}.
 \label{eq:app-periodic-explicit-twistors}
\end{align}
The remaining twistors are obtained by the period transformation
\begin{equation}
 Z_{i+4}=\mathcal P Z_i,
 \qquad
 \mathcal P=
 \begin{pmatrix}
 1&0&0&0\\
 0&1&0&0\\
 1&0&1&0\\
 0&1&0&1
 \end{pmatrix}.
 \label{eq:app-periodic-monodromy}
\end{equation}
Eqs.~\eqref{eq:app-periodic-boundary-measurement}--
\eqref{eq:app-periodic-monodromy} thus construct the periodic
momentum twistor configuration directly from the five positive face
variables.  Direct substitution of these twistors into
Eq.~\eqref{eq:g48-positive-x-chart} recovers the five face variables
$x_1,\ldots,x_5$ exactly.

\paragraph{Relation between the parametrizations.}
The change of variables is obtained by substituting the OPE twistors into
Eq.~\eqref{eq:g48-positive-x-chart}.  Define
\begin{equation}
 \begin{aligned}
 D&=1+S_1^2+T_1^2+T_2^2,\\
 P&=S_1^2(1+T_2^2)+\varphi S_2T_2(S_1^2+T_1^2),
 \qquad Q=\varphi S_1^2(1+T_2^2)+S_2T_2(S_1^2+T_1^2),\\
 R&=\varphi S_1^2\bigl[(1+T_1^2+T_2^2)^2+S_1^2T_1^2\bigr]+S_1^2S_2T_2(1+\varphi^2)(1+T_2^2)+\varphi S_2^2T_2^2(S_1^2+T_1^2).
 \end{aligned}
 \label{eq:app-ope-x-auxiliary}
\end{equation}
The five positive coordinates are then
\begin{equation}
 \begin{aligned}
 x_1&=\frac{PQ}{(1+T_2^2)R},&
 x_2&=\frac{(1+T_2^2)R}{S_2T_1^2T_2D^2},\\
 x_3&=\frac{\varphi S_1^2T_1^2D^2}{PQ},&
 x_4&=\frac{(1+T_2^2)R}{\varphi^2S_2T_1^2T_2D^2},&
 x_5&=\frac{\varphi S_2^2D^2}{PQ}.
 \end{aligned}
 \label{eq:app-ope-to-positive-x}
\end{equation}
In particular,
\begin{equation}
 \frac{x_2}{x_4}=\varphi^2,
 \qquad
 x_1x_2x_3=\frac{\varphi S_1^2}{S_2T_2},
 \qquad
 x_1x_2x_5=\frac{\varphi S_2}{T_1^2T_2}.
 \label{eq:app-ope-x-simple-relations}
\end{equation}
Thus the map from the OPE variables to the positive coordinates is rational.
The inverse map requires a choice of square-root branch, beginning with
$\varphi=\sqrt{x_2/x_4}$; in the positive region we take the positive branch.

\bibliographystyle{JHEP}
\bibliography{references}

\end{document}

%% file: figures/amplitudes_form_factors_roadmap.tex
\pgfdeclarehorizontalshading{periodfadewestmask}{100bp}{
  color(0bp)=(pgftransparent!100);
  color(45bp)=(pgftransparent!60);
  color(75bp)=(pgftransparent!0);
  color(100bp)=(pgftransparent!0)}
\pgfdeclarehorizontalshading{periodfadeeastmask}{100bp}{
  color(0bp)=(pgftransparent!0);
  color(25bp)=(pgftransparent!0);
  color(55bp)=(pgftransparent!60);
  color(100bp)=(pgftransparent!100)}
\pgfdeclarefading{period fade west}{%
  \pgfuseshading{periodfadewestmask}}
\pgfdeclarefading{period fade east}{%
  \pgfuseshading{periodfadeeastmask}}
\definecolor{roadmapedge}{HTML}{355F7C}
\definecolor{roadmaplimit}{HTML}{73797E}
\definecolor{roadmapanti}{HTML}{D28A55}
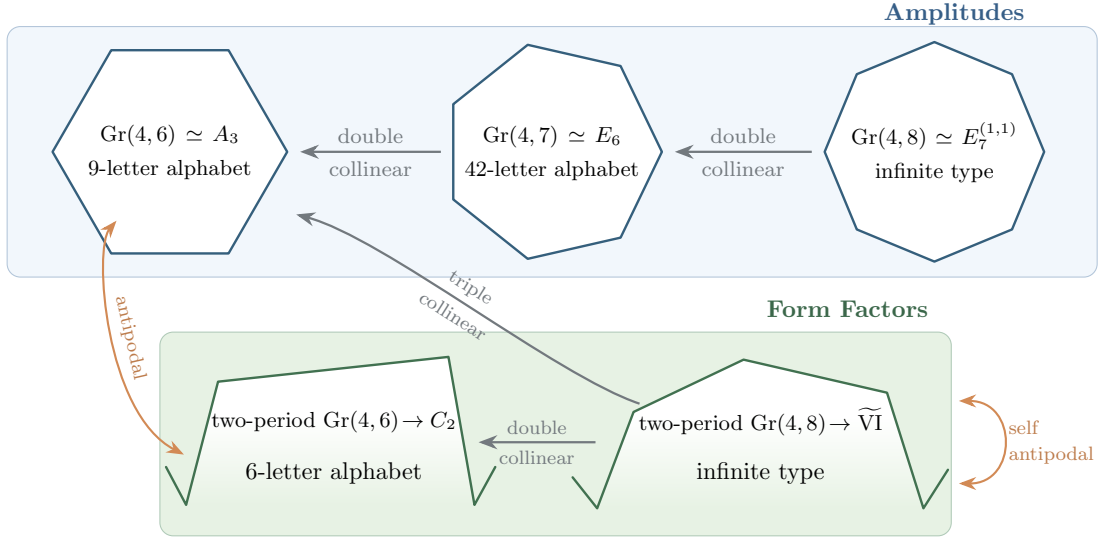
\begin{figure}[t]
\centering
\begin{tikzpicture}[
  x=1cm,y=1cm,scale=0.91,transform shape,
  amptext/.style={align=center,text width=27mm,font=\footnotesize,inner sep=0pt},
  amppolygon/.style={draw=roadmapedge,fill=white,line width=0.90pt,
    line join=round},
  periodcontour/.style={draw=accentgreen!82!black,line width=0.9pt,
    line cap=round,line join=round},
  periodgradient/.style={shade,top color=white,middle color=white,
    bottom color=periodone},
  periodlowergradient/.style={shade,top color=white,
    bottom color=periodone},
  limitarrow/.style={-{Stealth[length=2.5mm]},line width=0.85pt,
    draw=roadmaplimit},
  arrowlabel/.style={inner sep=1pt,font=\scriptsize,text=roadmaplimit}
]
  \path[draw=accentblue!40,fill=mutblue!38,rounded corners=5pt]
    (-2.87,0.18) rectangle (12.97,3.82);
  \path[draw=accentgreen!45,fill=periodone,rounded corners=5pt]
    (-0.65,-3.58) rectangle (10.85,-0.62);

  \node[font=\small\bfseries,text=accentblue!85!black,anchor=east]
    at (12.05,4.00) {Amplitudes};
  \node[font=\small\bfseries,text=accentgreen!75!black,anchor=east]
    at (10.65,-0.28) {Form Factors};

  \begin{scope}[cm={1.10,0,0,1.10,(-0.505,0.02)},transform shape]
  \begin{scope}[shift={(0,1.80)}]
    \path[amppolygon] (0:1.55)
      \foreach \a in {60,120,180,240,300} {--(\a:1.55)} --cycle;
  \end{scope}
  \begin{scope}[shift={(5.05,1.80)}]
    \path[amppolygon] (0:1.45)
      \foreach \a in {51.4286,102.8571,154.2857,205.7143,
        257.1429,308.5714} {--(\a:1.45)} --cycle;
  \end{scope}
  \begin{scope}[shift={(10.10,1.80)}]
    \path[amppolygon] (0:1.45)
      \foreach \a in {45,90,135,180,225,270,315} {--(\a:1.45)} --cycle;
  \end{scope}

  \draw[limitarrow] (8.48,1.80)--
    node[arrowlabel,pos=.50,above=2pt]{double}
    node[arrowlabel,pos=.50,below=2pt]{collinear}
    (6.67,1.80);
  \draw[limitarrow] (3.58,1.80)--
    node[arrowlabel,pos=.50,above=2pt]{double}
    node[arrowlabel,pos=.50,below=2pt]{collinear}
    (1.72,1.80);
  \end{scope}

  \node[amptext] (a6) at (-0.505,2.00)
    {$\Gr(4,6)\simeq A_3$\\[2pt]9-letter alphabet};
  \node[amptext] (a7) at (5.05,1.98)
    {$\Gr(4,7)\simeq E_6$\\[2pt]42-letter alphabet};
  \node[amptext,text width=29mm] (a8) at (10.605,2.00)
    {$\Gr(4,8)\simeq E_7^{(1,1)}$\\[2pt]infinite type};

  \begin{scope}[cm={1.08,0,0,1,(-0.1504,0)}]
  \coordinate (f3prev) at (-0.38,-2.58);
  \coordinate (f3z1) at (-0.11,-3.13);
  \coordinate (f3z2) at (0.32,-1.34);
  \coordinate (f3z3) at (3.42,-0.98);
  \coordinate (f3z1p) at (3.77,-3.13);
  \coordinate (f3next) at (4.05,-2.58);
  \path[periodgradient,path fading=period fade west,fit fading=true]
    (f3prev)--(f3z1)--(-0.38,-3.13)--cycle;
  \begin{scope}
    \clip (f3z1)--(f3z2)--(f3z3)--(f3z1p)--cycle;
    \path[periodlowergradient]
      (-0.11,-3.13) rectangle (3.77,-1.86);
    \path[fill=white]
      (-0.11,-1.86) rectangle (3.77,-0.90);
  \end{scope}
  \path[periodgradient,path fading=period fade east,fit fading=true]
    (f3z1p)--(f3next)--(4.05,-3.13)--cycle;
  \draw[periodcontour]
    (f3prev)--(f3z1)--(f3z2)--(f3z3)--(f3z1p)--(f3next);
  \end{scope}

  \begin{scope}[cm={1.08,0,0,1,(-0.6464,0)}]
  \coordinate (f4prev) at (5.55,-2.73);
  \coordinate (f4z1) at (5.88,-3.18);
  \coordinate (f4z2) at (6.37,-1.78);
  \coordinate (f4z3) at (7.85,-1.02);
  \coordinate (f4z4) at (9.78,-1.50);
  \coordinate (f4z1p) at (10.27,-3.18);
  \coordinate (f4next) at (10.60,-2.62);
  \path[periodgradient,path fading=period fade west,fit fading=true]
    (f4prev)--(f4z1)--(5.55,-3.18)--cycle;
  \begin{scope}
    \clip (f4z1)--(f4z2)--(f4z3)--(f4z4)--(f4z1p)--cycle;
    \path[periodlowergradient]
      (5.88,-3.18) rectangle (10.27,-1.86);
    \path[fill=white]
      (5.88,-1.86) rectangle (10.27,-0.90);
  \end{scope}
  \path[periodgradient,path fading=period fade east,fit fading=true]
    (f4z1p)--(f4next)--(10.60,-3.18)--cycle;
  \draw[periodcontour]
    (f4prev)--(f4z1)--(f4z2)--(f4z3)--(f4z4)--(f4z1p)--(f4next);
  \end{scope}

  \draw[limitarrow] (5.68,-2.22)--
    node[arrowlabel,pos=.50,above=2pt]{double}
    node[arrowlabel,pos=.50,below=2pt]{collinear}
    (3.98,-2.22);
  \draw[limitarrow] (6.32,-1.66)
    .. controls (5.10,-1.25) and (2.60,0.75) ..
    node[arrowlabel,pos=.52,above=2pt,sloped]{triple}
    node[arrowlabel,pos=.52,below=2pt,sloped]{collinear}
    (1.33,1.09);
  \draw[{Stealth[length=2.3mm]}-{Stealth[length=2.3mm]},thick,
    roadmapanti] (-1.28,1.00)
    .. controls (-1.62,0.60) and (-1.45,-1.80) ..
    node[inner sep=1pt,font=\scriptsize,pos=.53,above=2pt,sloped,
      text=roadmapanti!90!black]{antipodal}
    (-0.30,-2.40);
  \draw[{Stealth[length=2.3mm]}-{Stealth[length=2.3mm]},thick,roadmapanti]
    (10.96,-1.62)
    .. controls (11.82,-1.62) and (11.82,-2.82) ..
    (10.96,-2.82);
  \node[inner sep=0pt,font=\scriptsize,align=left,anchor=west,
    text=roadmapanti!90!black] at (11.68,-2.22)
    {self\\antipodal};

  \node[font=\footnotesize] at (1.88,-1.92)
    {two-period $\Gr(4,6)\!\to C_2$};
  \node[font=\small] at (1.88,-2.67) {6-letter alphabet};
  \node[font=\footnotesize] at (8.08,-1.92)
    {two-period $\Gr(4,8)\!\to\widetilde{\mathrm{VI}}$};
  \node[font=\small] at (8.08,-2.67) {infinite type};
\end{tikzpicture}
\caption{Cluster algebras for amplitude and form factor alphabets.  The arrows
indicate collinear limits and antipodal relations.}
\label{fig:amplitudes-form-factors-roadmap}
\end{figure}

%% file: figures/periodic_twistors_four_point_manuscript.tex
\begin{figure}[t]
 \centering
 \begin{tikzpicture}[
   x=1.65cm,
   y=1.18cm,
   >=Latex,
   contour/.style={line width=0.85pt,draw=black!80},
   twistorpoint/.style={circle,draw=black,fill=black,inner sep=1.35pt},
   zlabel/.style={font=\normalsize,inner sep=1pt},
   xlabel/.style={font=\scriptsize,fill=white,inner sep=1pt}
 ]
  \coordinate (x1)   at (0,0);
  \coordinate (x2)   at (0.72,1.18);
  \coordinate (x3)   at (1.55,1.48);
  \coordinate (x4)   at (2.35,0.86);
  \coordinate (x1p)  at (3.12,0);
  \coordinate (x2p)  at (3.84,1.18);
  \coordinate (x3p)  at (4.67,1.48);
  \coordinate (x4p)  at (5.47,0.86);
  \coordinate (x1pp) at (6.24,0);
  \coordinate (x4m)  at (-0.34,0.38);
  \coordinate (x2pp) at (6.58,0.56);

  \draw[contour]
   (x4m)--(x1)--(x2)--(x3)--(x4)--(x1p)--(x2p)--(x3p)--(x4p)--(x1pp)--(x2pp);
  \node at (-0.63,0.58) {$\cdots$};
  \node at (6.87,0.76) {$\cdots$};

  \foreach \p in {x1,x2,x3,x4,x1p,x2p,x3p,x4p,x1pp}
    \node[twistorpoint] at (\p) {};

  \node[zlabel,below=3pt of x1]   {$Z_1$};
  \node[zlabel,above left=2pt of x2] {$Z_2$};
  \node[zlabel,above=3pt of x3]   {$Z_3$};
  \node[zlabel,above right=2pt of x4] {$Z_4$};
  \node[zlabel,below=3pt of x1p]  {$Z_5$};
  \node[zlabel,above left=2pt of x2p] {$Z_6$};
  \node[zlabel,above=3pt of x3p]  {$Z_7$};
  \node[zlabel,above right=2pt of x4p] {$Z_8$};
  \node[zlabel,below=3pt of x1pp] {$Z_9$};

  \node[xlabel,below left=4pt] at ($(x4m)!0.52!(x1)$) {$x_1$};
  \node[xlabel,below right=1pt] at ($(x1)!0.50!(x2)$) {$x_2$};
  \node[xlabel,below=2pt] at ($(x2)!0.50!(x3)$) {$x_3$};
  \node[xlabel,below left=1pt] at ($(x3)!0.50!(x4)$) {$x_4$};
  \node[xlabel,below left=1pt] at ($(x4)!0.50!(x1p)$) {$x_5$};
  \node[xlabel,below right=1pt] at ($(x1p)!0.50!(x2p)$) {$x_6$};
  \node[xlabel,below=2pt] at ($(x2p)!0.50!(x3p)$) {$x_7$};
  \node[xlabel,below left=1pt] at ($(x3p)!0.50!(x4p)$) {$x_8$};
  \node[xlabel,below left=1pt] at ($(x4p)!0.50!(x1pp)$) {$x_9$};
 \end{tikzpicture}
 \caption{The periodic momentum twistor contour for a four-point form factor.
 Black vertices denote momentum twistors $Z_i$, while each segment
 $(Z_{i-1}Z_i)$ corresponds to the dual point $x_i$.  A period shift acts as
 $Z_i\mapsto Z_{i+4}=\cP Z_i$ and $x_i\mapsto x_{i+4}=x_i+q$.}
 \label{fig:intro-periodic-twistors}
\end{figure}
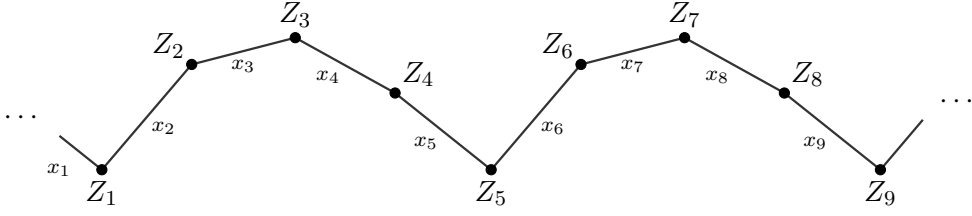

%% file: figures/g46_c2_seed.tex
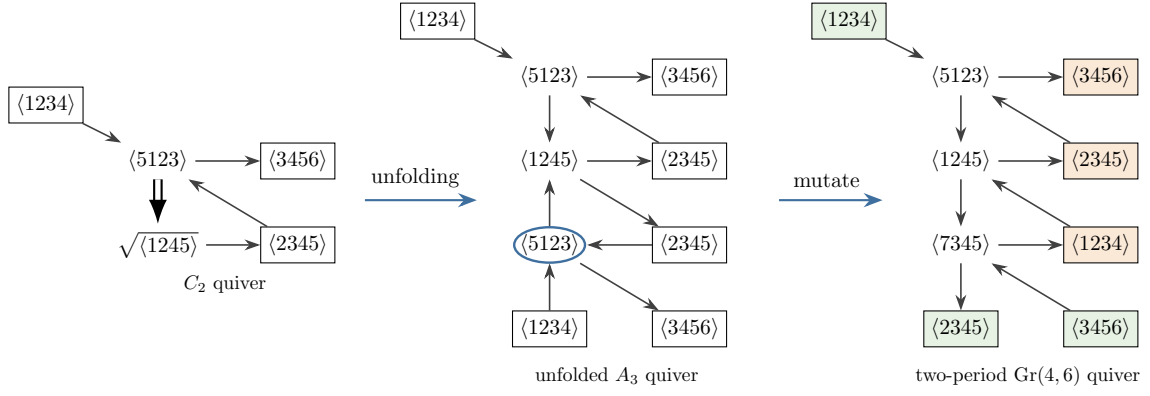
\begin{figure}[t]
\centering
\begin{tikzpicture}[scale=0.74,transform shape]
  \begin{scope}[xshift=-9.7cm]
    \begin{scope}[yshift=-1.5cm]
    \node (c1) at (0,0) {$\br{5123}$};
    \node (c2) at (0,-1.5) {$\sqrt{\br{1245}}$};
    \node[draw,rectangle] (cf1) at (-2,1) {$\br{1234}$};
    \node[draw,rectangle] (cf2) at (2.5,-1.5) {$\br{2345}$};
    \node[draw,rectangle] (cf3) at (2.5,0) {$\br{3456}$};
    \draw[double,double distance=2pt,line width=0.8pt,-Latex]
      (c1)--(c2);
    \draw[qarrow] (cf1)--(c1);
    \draw[qarrow] (c1)--(cf3);
    \draw[qarrow] (c2)--(cf2);
    \draw[qarrow] (cf2)--(c1);
    \end{scope}
    \node[font=\small] at (1.2,-3.7) {$C_2$ quiver};
  \end{scope}

  \draw[flowarrow] (-6.00,-2.2)-- (-4.00,-2.2);
  \node[above] at (-5.1,-2.1) {unfolding};

  \begin{scope}[xshift=-2.7cm]
    \node (m1) at (0,0) {$\br{5123}$};
    \node (m2) at (0,-1.5) {$\br{1245}$};
    \node (m3) at (0,-3) {$\br{5123}$};
    \node[draw,rectangle] (f1) at (-2,1) {$\br{1234}$};
    \node[draw,rectangle] (f2) at (0,-4.5) {$\br{1234}$};
    \node[draw,rectangle] (f3) at (2.5,0) {$\br{3456}$};
    \node[draw,rectangle] (f4) at (2.5,-1.5) {$\br{2345}$};
    \node[draw,rectangle] (f5) at (2.5,-3) {$\br{2345}$};
    \node[draw,rectangle] (f6) at (2.5,-4.5) {$\br{3456}$};
    \draw[qarrow] (m1)--(m2);
    \draw[qarrow] (m3)--(m2);
    \draw[qarrow] (f2)--(m3);
    \draw[qarrow] (f1)--(m1);
    \draw[qarrow] (m1)--(f3);
    \draw[qarrow] (m2)--(f4);
    \draw[qarrow] (f5)--(m3);
    \draw[qarrow] (f4)--(m1);
    \draw[qarrow] (m2)--(f5);
    \draw[qarrow] (m3)--(f6);
    \draw[accentblue,line width=0.9pt] (m3) ellipse [x radius=18pt,y radius=9pt];
    \node[font=\small] at (1.2,-5.35) {unfolded $A_3$ quiver};
  \end{scope}

   \draw[flowarrow] (1.40,-2.2)--(3.25,-2.2);
   \node[above] at (2.25,-2.1) {mutate};

  \begin{scope}[xshift=4.65cm]
    \node (n1) at (0,0) {$\br{5123}$};
    \node (n2) at (0,-1.5) {$\br{1245}$};
    \node (n3) at (0,-3) {$\br{7345}$};
    \node[draw,rectangle,fill=periodone] (g1) at (-2,1) {$\br{1234}$};
    \node[draw,rectangle,fill=periodone] (g2) at (0,-4.5) {$\br{2345}$};
    \node[draw,rectangle,fill=periodone] (g3) at (2.5,-4.5) {$\br{3456}$};
    \node[draw,rectangle,fill=periodtwo] (g4) at (2.5,-3) {$\br{1234}$};
    \node[draw,rectangle,fill=periodtwo] (g5) at (2.5,-1.5) {$\br{2345}$};
    \node[draw,rectangle,fill=periodtwo] (g6) at (2.5,0) {$\br{3456}$};
    \draw[qarrow] (n1)--(n2);
    \draw[qarrow] (n2)--(n3);
    \draw[qarrow] (n3)--(g2);
    \draw[qarrow] (g1)--(n1);
    \draw[qarrow] (n1)--(g6);
    \draw[qarrow] (n2)--(g5);
    \draw[qarrow] (n3)--(g4);
    \draw[qarrow] (g5)--(n1);
    \draw[qarrow] (g4)--(n2);
    \draw[qarrow] (g3)--(n3);
    \node[font=\small] at (1.2,-5.35) {two-period $\Gr(4,6)$ quiver};
  \end{scope}
\end{tikzpicture}
\caption{The three-point $C_2$ quiver on the left unfolds to the $A_3$
quiver in the middle.  Mutation at the
circled node produces the two-period $\Gr(4,6)$ quiver on the right; the two
colors distinguish the frozen coefficients associated with the two periods.}
\label{fig:g46-c2-seed}
\end{figure}

%% file: figures/g46_c2_exchange_hexagon.tex
\begin{figure}[t]
\centering
\begin{tikzpicture}[x=1cm,y=1cm,scale=0.85,transform shape]
  \tikzset{c2cluster/.style={draw=black!65,rounded corners=2pt,
      fill=black!2,inner xsep=5pt,inner ysep=4pt,font=\small}}
  \coordinate (h0) at (0, 2.75);
  \coordinate (h1) at (3.10, 1.1);
  \coordinate (h2) at (3.10,-1.1);
  \coordinate (h3) at (0,-2.75);
  \coordinate (h4) at (-3.10,-1.1);
  \coordinate (h5) at (-3.10, 1.1);

  \draw[semithick,black!65]
    (h0)--(h1)--(h2)--(h3)--(h4)--(h5)--cycle;
  \node[c2cluster] at (h0)
    {$\displaystyle\left\{\frac{vw}{u},\frac{1-w}{w}\right\}$};
  \node[c2cluster] at (h1)
    {$\displaystyle\left\{\frac{u}{vw},\frac{v}{1-v}\right\}$};
  \node[c2cluster] at (h2)
    {$\displaystyle\left\{\frac{uv}{w},\frac{1-v}{v}\right\}$};
  \node[c2cluster] at (h3)
    {$\displaystyle\left\{\frac{w}{uv},\frac{u}{1-u}\right\}$};
  \node[c2cluster] at (h4)
    {$\displaystyle\left\{\frac{uw}{v},\frac{1-u}{u}\right\}$};
  \node[c2cluster] at (h5)
    {$\displaystyle\left\{\frac{v}{uw},\frac{w}{1-w}\right\}$};

  \node[font=\scriptsize,fill=white,inner sep=1pt] at (1.6,1.9) {$\mu_1$};
  \node[font=\scriptsize,fill=white,inner sep=1pt] at (3.10,0) {$\mu_2$};
  \node[font=\scriptsize,fill=white,inner sep=1pt] at (1.6,-1.9) {$\mu_1$};
  \node[font=\scriptsize,fill=white,inner sep=1pt] at (-1.6,-1.9) {$\mu_2$};
  \node[font=\scriptsize,fill=white,inner sep=1pt] at (-3.10,0) {$\mu_1$};
  \node[font=\scriptsize,fill=white,inner sep=1pt] at (-1.6,1.9) {$\mu_2$};
  \node[font=\large] at (0,0) {$C_2$};
\end{tikzpicture}
\caption{The $C_2$ exchange graph associated with the three-point form factor
alphabet.  Its six vertices represent the six clusters and are labelled by
the corresponding $\mathcal X$-coordinates in terms of $\{u,v,w\}$; the edge
labels indicate the mutations relating adjacent clusters.}
\label{fig:g46-c2-exchange-hexagon}
\end{figure}
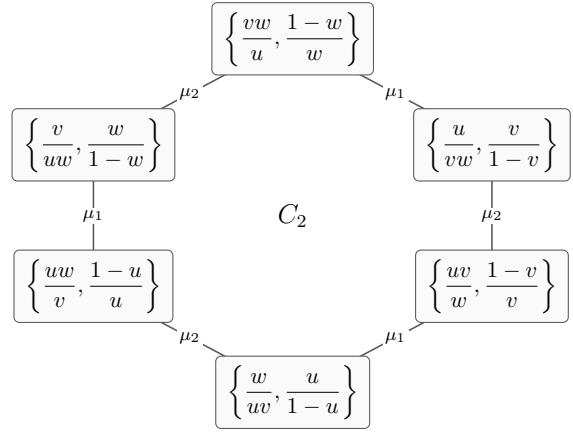

%% file: figures/g48_periodicized_extended_seed.tex
\begin{figure}[t]
\centering
\begin{tikzpicture}[scale=0.85,transform shape]
  \node (m1) at (0,0) {$1:\br{1235}$};
  \node (m2) at (2.5,0) {$2:\br{1236}$};
  \node (m3) at (5,0) {$3:\br{1237}$};
  \node (m4) at (0,-1.5) {$4:\br{1245}$};
  \node (m5) at (2.5,-1.5) {$5:\br{1256}$};
  \node (m6) at (5,-1.5) {$6:\br{2356}$};
  \node (m7) at (0,-3) {$7:\br{1345}$};
  \node (m8) at (2.5,-3) {$8:\br{1456}$};
  \node (m9) at (5,-3) {$9:\br{1567}$};

  \node[draw,rectangle,fill=periodone] (f10) at (-2,1)
    {$\br{1234}$};
  \node[draw,rectangle,fill=periodone] (f17) at (0,-4.5)
    {$\br{2345}$};
  \node[draw,rectangle,fill=periodone] (f16) at (2.5,-4.5)
    {$\br{3456}$};
  \node[draw,rectangle,fill=periodone] (f15) at (5,-4.5)
    {$\br{4567}$};
  \node[draw,rectangle,fill=periodtwo] (f14) at (7.5,-4.5)
    {$\br{1234}$};
  \node[draw,rectangle,fill=periodtwo] (f13) at (7.5,-3)
    {$\br{2345}$};
  \node[draw,rectangle,fill=periodtwo] (f12) at (7.5,-1.5)
    {$\br{3456}$};
  \node[draw,rectangle,fill=periodtwo] (f11) at (7.5,0)
    {$\br{4567}$};

  \draw[qarrow] (m1)--(m2); \draw[qarrow] (m1)--(m4);
  \draw[qarrow] (m2)--(m3); \draw[qarrow] (m2)--(m5);
  \draw[qarrow] (m3)--(m6); \draw[qarrow] (m4)--(m5);
  \draw[qarrow] (m4)--(m7); \draw[qarrow] (m5)--(m6);
  \draw[qarrow] (m5)--(m8); \draw[qarrow] (m6)--(m9);
  \draw[qarrow] (m7)--(m8); \draw[qarrow] (m8)--(m9);
  \draw[qarrow] (m5)--(m1); \draw[qarrow] (m6)--(m2);
  \draw[qarrow] (m8)--(m4); \draw[qarrow] (m9)--(m5);

  \draw[qarrow,shorten <=2pt] (f10)--(m1);
  \draw[qarrow,shorten >=2pt] (m7)--(f17);
  \draw[qarrow,shorten <=2pt] (f16)--(m7);
  \draw[qarrow,shorten >=2pt] (m8)--(f16);
  \draw[qarrow,shorten <=2pt] (f15)--(m8);
  \draw[qarrow,shorten >=2pt] (m9)--(f15);
  \draw[qarrow,shorten <=2pt] (f14)--(m9);
  \draw[qarrow,shorten <=2pt] (f13)--(m6);
  \draw[qarrow,shorten >=2pt] (m9)--(f13);
  \draw[qarrow,shorten <=2pt] (f12)--(m3);
  \draw[qarrow,shorten >=2pt] (m6)--(f12);
  \draw[qarrow,shorten >=2pt] (m3)--(f11);
\end{tikzpicture}
\caption{A two-period $\Gr(4,8)$ quiver whose coordinates are evaluated on
four-point periodic momentum twistor kinematics.  Mutable nodes are labelled by their $\mathcal A$-coordinates.
Boxed nodes are frozen; the two colors indicate the two periods.}
\label{fig:g48-extended-seed}
\end{figure}
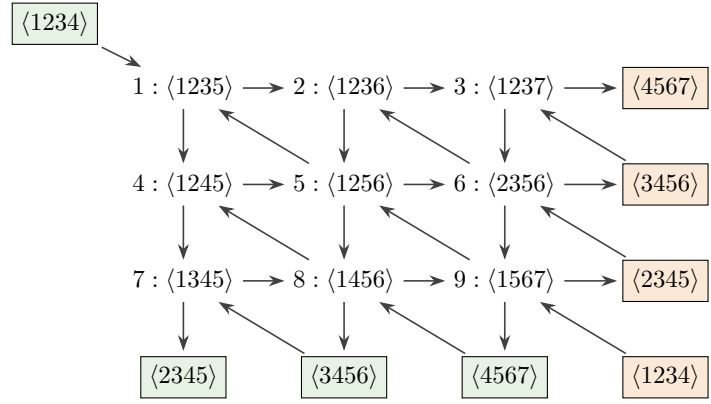

%% file: figures/g48_foldable_extended_quiver.tex
\begin{figure}[t]
\centering
\begin{tikzpicture}[scale=0.85,transform shape]
  \node (m1) at (0,0) {$1:\br{1235}$};
  \node (m2) at (2.5,0) {$2:\br{1236}$};
  \node (m3) at (5,0) {$3:\br{2456}$};
  \node (m4) at (0,-1.5) {$4:\br{1245}$};
  \node (m5) at (2.5,-1.5) {$5:\br{1256}$};
  \node (m6) at (5,-1.5) {$6:\br{1245}$};
  \node (m7) at (0,-3) {$7:\br{2456}$};
  \node (m8) at (2.5,-3) {$8:\br{1236}$};
  \node (m9) at (5,-3) {$9:\br{1235}$};

  \node[draw,rectangle,fill=periodone] (f10) at (-2,1.5)
    {$\br{1234}$};
  \node[draw,rectangle,fill=periodone] (f17) at (-2,-2.25)
    {$\br{2345}$};
  \node[draw,rectangle,fill=periodone] (f16) at (-2,-4.5)
    {$\br{3456}$};
  \node[draw,rectangle,fill=periodone] (f15) at (1.25,-4.5)
    {$\br{4567}$};
  \node[draw,rectangle,fill=periodtwo] (f14) at (7,-4.5)
    {$\br{1234}$};
  \node[draw,rectangle,fill=periodtwo] (f13) at (7,-0.75)
    {$\br{2345}$};
  \node[draw,rectangle,fill=periodtwo] (f12) at (7,1.5)
    {$\br{3456}$};
  \node[draw,rectangle,fill=periodtwo] (f11) at (3.75,1.5)
    {$\br{4567}$};

  \draw[qarrow] (m1)--(m2); \draw[qarrow] (m3)--(m2);
  \draw[qarrow] (m4)--(m5); \draw[qarrow] (m6)--(m5);
  \draw[qarrow] (m7)--(m8); \draw[qarrow] (m9)--(m8);
  \draw[qarrow] (m1)--(m4); \draw[qarrow] (m7)--(m4);
  \draw[qarrow] (m2)--(m5); \draw[qarrow] (m8)--(m5);
  \draw[qarrow] (m3)--(m6); \draw[qarrow] (m9)--(m6);
  \draw[qarrow] (m5)--(m1); \draw[qarrow] (m5)--(m3);
  \draw[qarrow] (m5)--(m7); \draw[qarrow] (m5)--(m9);

  \draw[qarrow,shorten <=2pt] (f10)--(m1);
  \draw[qarrow,shorten >=2pt] (m4)--(f17);
  \draw[qarrow,shorten <=2pt] (f17)--(m7);
  \draw[qarrow,shorten >=2pt] (m7)--(f16);
  \draw[qarrow,shorten <=2pt] (f15)--(m7);
  \draw[qarrow,shorten >=2pt] (m8)--(f15);
  \draw[qarrow,shorten <=2pt] (f14)--(m9);
  \draw[qarrow,shorten >=2pt] (m6)--(f13);
  \draw[qarrow,shorten <=2pt] (f13)--(m3);
  \draw[qarrow,shorten >=2pt] (m3)--(f12);
  \draw[qarrow,shorten <=2pt] (f11)--(m3);
  \draw[qarrow,shorten >=2pt] (m2)--(f11);
\end{tikzpicture}
\caption{The quiver obtained from Fig.~\ref{fig:g48-extended-seed} by
successively mutating nodes $3$, $6$, $7$ and $8$.  It is symmetric under
$1\leftrightarrow9$, $2\leftrightarrow8$, $3\leftrightarrow7$ and
$4\leftrightarrow6$, with node $5$ fixed.  The symmetry also exchanges the
boxed frozen nodes between the two colored periods.}
\label{fig:g48-foldable-extended-quiver}
\end{figure}
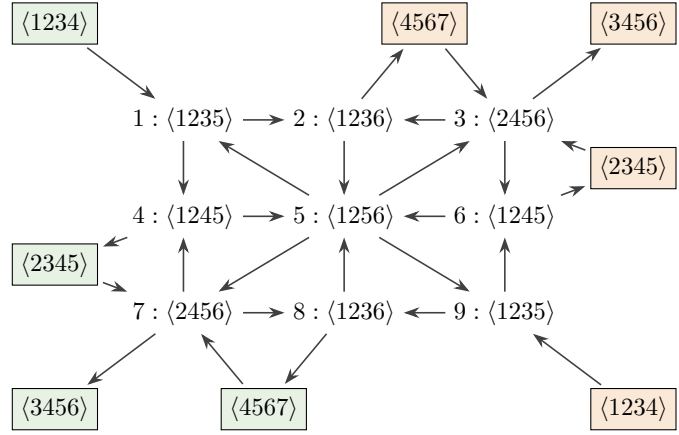

%% file: figures/g48_folded_quiver.tex
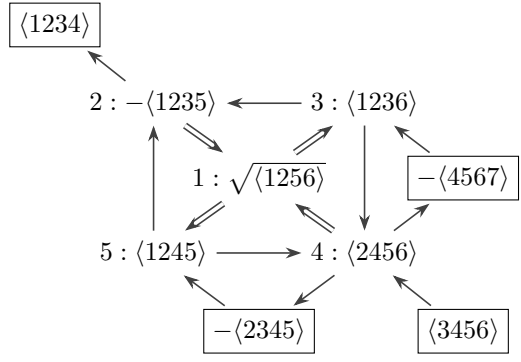
\begin{figure}[t]
\centering
\begin{tikzpicture}[scale=0.90,transform shape]
  \node (a1) at (0,0) {$1:\sqrt{\br{1256}}$};
  \node (a2) at (-1.55,1.10) {$2:-\br{1235}$};
  \node (a3) at (1.55,1.10) {$3:\br{1236}$};
  \node (a5) at (-1.55,-1.10) {$5:\br{1245}$};
  \node (a4) at (1.55,-1.10) {$4:\br{2456}$};

  \node[draw,rectangle] (f1) at (-3,2.25) {$\br{1234}$};
  \node[draw,rectangle] (f4) at (0,-2.25) {$-\br{2345}$};
  \node[draw,rectangle] (f3) at (3,-2.25) {$\br{3456}$};
  \node[draw,rectangle] (f2) at (3,0) {$-\br{4567}$};

  \draw[qarrow] (a3)--(a2);
  \draw[qarrow] (a5)--(a4);
  \draw[qarrow] (a5)--(a2);
  \draw[qarrow] (a3)--(a4);
  \draw[qarrow,double distance=1.4pt] (a2)--(a1);
  \draw[qarrow,double distance=1.4pt] (a1)--(a3);
  \draw[qarrow,double distance=1.4pt] (a1)--(a5);
  \draw[qarrow,double distance=1.4pt] (a4)--(a1);

  \draw[qarrow,shorten >=2pt] (a2)--(f1);
  \draw[qarrow,shorten <=2pt] (f4)--(a5);
  \draw[qarrow,shorten >=2pt] (a4)--(f4);
  \draw[qarrow,shorten <=2pt] (f3)--(a4);
  \draw[qarrow,shorten <=2pt] (f2)--(a3);
  \draw[qarrow,shorten >=2pt] (a4)--(f2);
\end{tikzpicture}
\caption{The five-node $\widetilde{\mathrm{VI}}$ quiver defined by the
extended exchange matrix in Eq.~\eqref{eq:g48-vi-extended-matrix} and the
$\mathcal A$-coordinates in Eq.~\eqref{eq:g48-lifted-coordinates}. Boxed nodes
are frozen, and double arrows indicate exchange-matrix entries of magnitude
$2$.}
\label{fig:g48-folded-quiver}
\end{figure}

%% file: figures/affine_a11_rooted_origin.tex
\begin{figure}[t]
\centering
\begin{tikzpicture}[scale=0.90,transform shape]
  \node (w) at (-2.15,0) {$w_0$};
  \node (z) at ( 2.15,0) {$z_0$};
  \node (b) at (0,1.70) {$b$};
  \node[draw,rectangle] (fw) at (-2.15,-1.45) {$f_w$};
  \node[draw,rectangle] (fz) at ( 2.15,-1.45) {$f_z$};

  \draw[-{Stealth[length=2.4mm]},thick,accentblue]
    ([yshift=4pt]w.east) to[bend left=15] ([yshift=4pt]z.west);
  \draw[-{Stealth[length=2.4mm]},thick,accentblue]
    ([yshift=-4pt]w.east) to[bend right=15] ([yshift=-4pt]z.west);
  \node[font=\scriptsize,accentblue,fill=white,inner sep=1.5pt]
    at (0,0) {$A_1^{(1)}$};

  \draw[qarrow]
    ([xshift=-3pt]b.south west)--([xshift=2pt,yshift=5pt]w.north east);
  \draw[qarrow]
    ([xshift=-2pt,yshift=5pt]z.north west)--([xshift=3pt]b.south east);
  \draw[qarrow,draw=gray!65,shorten >=2pt] (w)--(fw);
  \draw[qarrow,draw=gray!65,shorten <=2pt] (fz)--(z);
\end{tikzpicture}
\caption{A schematic $A_1^{(1)}$ configuration.  The remaining mutable
coordinates contribute the common monomial $b$, while the boxed quantities
$f_w$ and $f_z$ collect the frozen-coordinate contributions.}
\label{fig:a11-rooted-origin}
\end{figure}
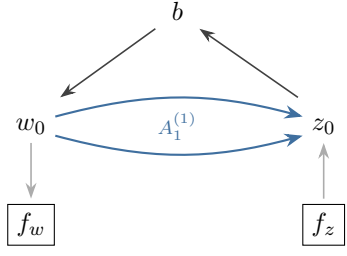

%% file: figures/affine_a11_mu21_quiver.tex
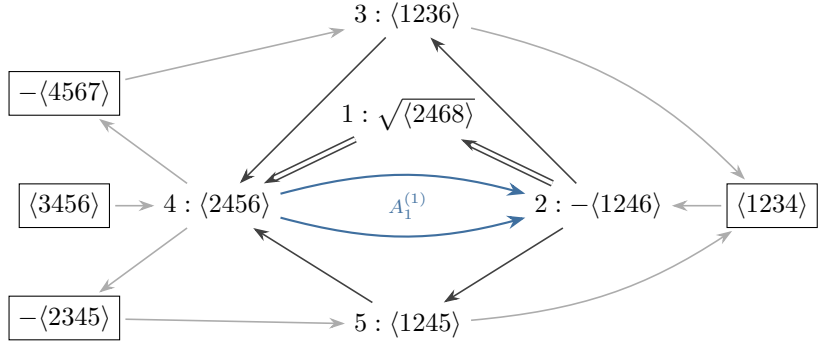
\begin{figure}[t]
\centering
\begin{tikzpicture}[scale=0.90,transform shape]
  \node (a4) at (-2.80,0) {$4:\br{2456}$};
  \node (a2) at ( 2.80,0) {$2:-\br{1246}$};
  \node (a1) at (0,1.35) {$1:\sqrt{\br{2468}}$};
  \node (a3) at (0,2.80) {$3:\br{1236}$};
  \node (a5) at (0,-1.75) {$5:\br{1245}$};

  \node[draw,rectangle] (f1) at (5.35,0)
    {$\br{1234}$};
  \node[draw,rectangle] (f4) at (-5.05,1.65)
    {$-\br{4567}$};
  \node[draw,rectangle] (f3) at (-5.05,0)
    {$\br{3456}$};
  \node[draw,rectangle] (f2) at (-5.05,-1.65)
    {$-\br{2345}$};

  \draw[qarrow,double distance=1.4pt] (a2)--(a1);
  \draw[qarrow,double distance=1.4pt] (a1)--(a4);
  \draw[qarrow] (a2)--(a3);
  \draw[qarrow] (a3)--(a4);
  \draw[qarrow] (a2)--(a5);
  \draw[qarrow] (a5)--(a4);

  \draw[-{Stealth[length=2.4mm]},thick,accentblue]
    ([yshift=5pt]a4.east) to[bend left=15]
    ([yshift=5pt]a2.west);
  \draw[-{Stealth[length=2.4mm]},thick,accentblue]
    ([yshift=-5pt]a4.east) to[bend right=15]
    ([yshift=-5pt]a2.west);
  \node[font=\scriptsize,accentblue,fill=white,inner sep=1.5pt]
    at (0,0) {$A_1^{(1)}$};

  \draw[qarrow,draw=gray!65,shorten <=2pt] (f1)--(a2);
  \draw[qarrow,draw=gray!65,shorten >=2pt]
    (a3) to[bend left=16] (f1);
  \draw[qarrow,draw=gray!65,shorten >=2pt]
    (a5) to[bend right=14] (f1);
  \draw[qarrow,draw=gray!65,shorten <=2pt] (f4)--(a3);
  \draw[qarrow,draw=gray!65,shorten >=2pt] (a4)--(f4);
  \draw[qarrow,draw=gray!65,shorten <=2pt] (f3)--(a4);
  \draw[qarrow,draw=gray!65,shorten >=2pt] (a4)--(f2);
  \draw[qarrow,draw=gray!65,shorten <=2pt] (f2)--(a5);
\end{tikzpicture}
\caption{The $\widetilde{\mathrm{VI}}$ quiver reached from the initial seed by
the mutation sequence $\{\mu_2,\mu_1\}$.  Nodes $4$ and $2$, identified with
$w_0$ and $z_0$, form the $A_1^{(1)}$ subquiver.}
\label{fig:affine-a11-mu21-quiver}
\end{figure}

%% file: figures/antipodal_folded_quiver.tex
\begin{figure}[t]
\centering
\begin{tikzpicture}[scale=0.92,transform shape,baseline=(current bounding box.center)]
  \node (a1) at (0,1.55) {$\bar x_1$};
  \node (a2) at (0,-1.55) {$\bar x_2$};
  \node (a3) at (-2.55,0) {$\bar x_3$};
  \node (a4) at (2.55,0) {$\bar x_4$};

  \draw[qarrow]
    ([xshift=-4pt]a2.north) to[bend left=7]
    ([xshift=-4pt]a1.south);
  \draw[qarrow]
    ([xshift=4pt]a2.north) to[bend right=7]
    ([xshift=4pt]a1.south);

  \draw[double,double distance=2pt,line width=0.8pt,-Latex] (a1)--(a3);
  \draw[double,double distance=2pt,line width=0.8pt,-Latex] (a1)--(a4);
  \draw[double,double distance=2pt,line width=0.8pt,-Latex] (a3)--(a2);
  \draw[double,double distance=2pt,line width=0.8pt,-Latex] (a4)--(a2);
\end{tikzpicture}
\hspace{1.2cm}
$
\overline B_{\rm p}=
\begin{pmatrix}
 0&-2& 2& 2\\
 2& 0&-2&-2\\
-1& 1& 0& 0\\
-1& 1& 0& 0
\end{pmatrix}
$
\caption{The folded quiver on the parity-preserving surface and its
principal exchange matrix, written in the coordinate order
$\{\bar x_1,\bar x_2,\bar x_3,\bar x_4\}$.  The two parallel arrows from
$\bar x_2$ to $\bar x_1$ represent two ordinary arrows, while the remaining
double-line arrows denote $C_2$-type valued edges.}
\label{fig:antipodal-folded-quiver}
\end{figure}
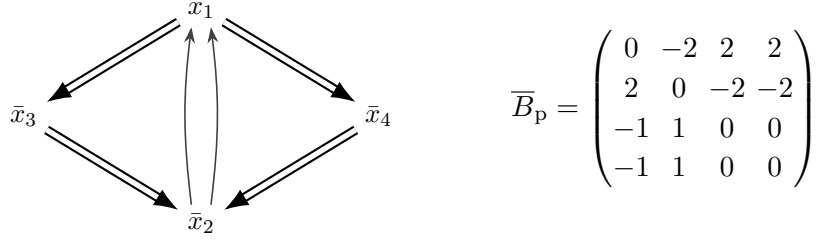

%% file: figures/triple_collinear_mutated_quiver.tex
\begin{figure}[t]
\centering
\begin{tikzpicture}[scale=0.68,transform shape]
  \filldraw[draw=accentblue!55,fill=mutblue!60,line width=0.5pt,
    rounded corners=5pt]
    (2.72,2.10) rectangle (3.58,-2.10);
  \begin{scope}[shift={(-1.05,0)}]
    \filldraw[draw=accentorange!65,fill=periodtwo!80,line width=0.5pt,
      rounded corners=5pt]
      (-1.66,-0.43) rectangle (1.66,0.43);
  \end{scope}

  \node[font=\small,accentorange] at (-1.10,0.92) {$C_2$};
  \node[font=\small,accentblue] at (3.96,-1.96) {$A_3$};

  \node (n3) at (-2.45,0) {$3$};
  \node (n1) at (0.35,0) {$1$};
  \node (n2) at (3.15,1.75) {$2$};
  \node (n5) at (3.15,0) {$5$};
  \node (n4) at (3.15,-1.75) {$4$};

  \draw[double,double distance=1.7pt,line width=0.7pt,-Latex]
    (n1.west)--(n3.east);
  \draw[double,double distance=1.7pt,line width=0.7pt,-Latex]
    (n5.west)--(n1.east);

  \draw[qarrow]
    ([xshift=1pt,yshift=14pt]n3.east)
    .. controls (-0.85,2.38) and (1.55,2.38) ..
    (n2.west);
  \draw[qarrow] (n2.south)--(n5.north);
  \draw[qarrow]
    ([xshift=1pt,yshift=-18pt]n3.east)
    .. controls (-0.85,-2.38) and (1.55,-2.38) ..
    (n4.west);
  \draw[qarrow] (n4.north)--(n5.south);

  \draw[-{Latex[length=3.0mm,width=2.2mm]},line width=1.0pt,accentblue]
    (4.35,0)--(5.30,0)
    node[midway,above=3pt,font=\small] {$2\sim4$}
    node[midway,below=4pt,font=\scriptsize] {folding};

  \begin{scope}[shift={(0.45,0)}]
  \begin{scope}[shift={(8.23,-0.92)},rotate=45]
    \filldraw[draw=accentblue!55,fill=mutblue!60,line width=0.5pt,
      rounded corners=4pt]
      (-1.52,-0.42) rectangle (1.52,0.42);
  \end{scope}
  \begin{scope}[shift={(6.60,0.85)},rotate=45]
    \filldraw[draw=accentorange!65,fill=periodtwo!80,line width=0.5pt,
      rounded corners=4pt]
      (-1.52,-0.42) rectangle (1.52,0.42);
  \end{scope}

  \node[font=\small,accentorange] at (5.85,1.68) {$C_2$};
  \node[font=\small,accentblue] at (8.98,-1.72) {$C_2$};

  \node (f3)  at (5.75,0) {$3$};
  \node (f1)  at (7.45,1.70) {$1$};
  \coordinate (f24) at (7.45,-1.70);
  \node[font=\scriptsize] at (7.45,-1.78) {$\{2,4\}$};
  \node (f5)  at (9.15,0) {$5$};

  \draw[double,double distance=1.7pt,line width=0.7pt,-Latex,
    shorten <=5pt,shorten >=5pt]
    (f1.center)--(f3.center);
  \draw[double,double distance=1.7pt,line width=0.7pt,-Latex,
    shorten <=5pt,shorten >=5pt]
    (f5.center)--(f1.center);
  \draw[double,double distance=1.7pt,line width=0.7pt,-Latex,
    shorten <=5pt,shorten >=5pt]
    (f3.center)--(f24.center);
  \draw[double,double distance=1.7pt,line width=0.7pt,-Latex,
    shorten <=5pt,shorten >=5pt]
    (f24.center)--(f5.center);

  \end{scope}

  \draw[-{Latex[length=3.0mm,width=2.2mm]},line width=1.0pt,accentblue]
    (10.30,0)--(11.45,0)
    node[midway,above=3pt,font=\small] {$g$}
    node[midway,below=4pt,font=\scriptsize]
      {antipodal};

  \begin{scope}[shift={(0.90,0)}]
  \begin{scope}[shift={(12.17,0.92)},rotate=45]
    \filldraw[draw=accentblue!55,fill=mutblue!60,line width=0.5pt,
      rounded corners=4pt]
      (-1.52,-0.42) rectangle (1.52,0.42);
  \end{scope}
  \begin{scope}[shift={(13.80,-0.85)},rotate=45]
    \filldraw[draw=accentorange!65,fill=periodtwo!80,line width=0.5pt,
      rounded corners=4pt]
      (-1.52,-0.42) rectangle (1.52,0.42);
  \end{scope}

  \node[font=\small,accentblue] at (11.42,1.67) {$C_2$};
  \node[font=\small,accentorange] at (14.55,-1.60) {$C_2$};

  \node (a3)  at (11.25,0) {$5$};
  \coordinate (a1) at (12.95,1.70);
  \node[font=\scriptsize] at (12.95,1.78) {$\{2,4\}$};
  \node (a24) at (12.95,-1.70) {$1$};
  \node (a5)  at (14.65,0) {$3$};

  \draw[double,double distance=1.7pt,line width=0.7pt,-Latex,
    shorten <=5pt,shorten >=5pt]
    (a1.center)--(a3.center);
  \draw[double,double distance=1.7pt,line width=0.7pt,-Latex,
    shorten <=5pt,shorten >=5pt]
    (a5.center)--(a1.center);
  \draw[double,double distance=1.7pt,line width=0.7pt,-Latex,
    shorten <=5pt,shorten >=5pt]
    (a3.center)--(a24.center);
  \draw[double,double distance=1.7pt,line width=0.7pt,-Latex,
    shorten <=5pt,shorten >=5pt]
    (a24.center)--(a5.center);

  \end{scope}

\end{tikzpicture}
\caption{The $\widetilde{\mathrm{VI}}$ quiver after successively mutating
nodes $3$, $1$, $2$ and $4$, and its collinear boundaries.  In the left panel, the blue and
orange regions indicate the $A_3$ triple-collinear and $C_2$ double-collinear
boundary subquivers, respectively.  On the parity-preserving surface,
identifying nodes $2$ and $4$ folds the $A_3$ subquiver to $C_2$, as shown in
the middle panel.  The antipodal cluster transformation exchanges the two
$C_2$ boundaries, giving the quiver on the right.}
\label{fig:triple-collinear-mutated-quiver}
\end{figure}
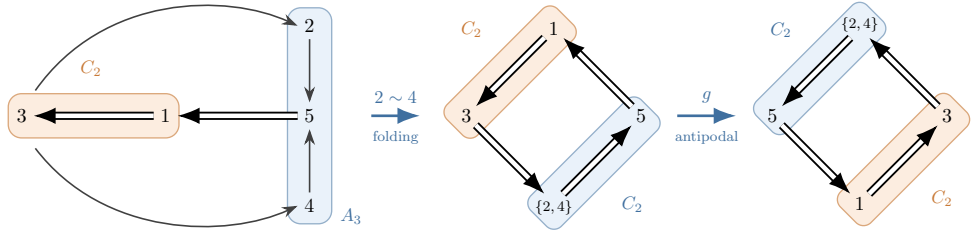

%% file: figures/periodic_plabic_graphs.tex

\begin{figure}[h]
\centering
\begin{minipage}[c]{0.49\textwidth}
\centering
\begin{tikzpicture}[
  scale=1.02,
  plabicedge/.style={draw=black,line width=1.15pt,line cap=round},
  plblack/.style={circle,draw=black,line width=0.9pt,fill={rgb,255:red,143;green,137;blue,204},
    minimum size=3.0mm,inner sep=0pt},
  plwhite/.style={circle,draw=black,line width=0.95pt,fill=white,
    minimum size=3.0mm,inner sep=0pt},
  orientation/.style={draw=black,line width=0.82pt,
    -{Stealth[length=1.75mm,width=1.25mm]}},
  annularcut/.style={draw=red!85!orange,line width=1.00pt,
    dash pattern=on 4.0pt off 2.3pt,line cap=round,
    preaction={draw=white,line width=2.35pt,line cap=round}},
  faceback/.style={line width=0.90pt,line join=round,line cap=round},
  facexone/.style={faceback,draw={rgb,255:red,184;green,116;blue,129},
    fill={rgb,255:red,249;green,230;blue,233}},
  facextwo/.style={faceback,draw={rgb,255:red,102;green,143;blue,183},
    fill={rgb,255:red,226;green,238;blue,249}},
  facexthree/.style={faceback,draw={rgb,255:red,105;green,153;blue,118},
    fill={rgb,255:red,230;green,244;blue,232}},
  facexfour/.style={faceback,draw={rgb,255:red,191;green,137;blue,82},
    fill={rgb,255:red,250;green,237;blue,220}},
  facexfive/.style={faceback,draw={rgb,255:red,139;green,116;blue,177},
    fill={rgb,255:red,239;green,232;blue,248}},
  facelabel/.style={font=\small,inner sep=0.5pt},
  xlabel/.style={font=\small,inner sep=0.5pt}
]
  \begin{scope}[rotate=22.5]

  \coordinate (W0) at (1.280,1.530);
  \coordinate (W1) at (0.153,1.075);
  \coordinate (W2) at (0.902,0.098);
  \coordinate (W3) at (2.179,-0.236);
  \coordinate (W4) at (-0.153,-1.075);
  \coordinate (W5) at (0.902,-1.512);
  \coordinate (W6) at (-0.902,1.512);
  \coordinate (W7) at (-0.902,-0.098);
  \coordinate (W8) at (-2.179,0.236);
  \coordinate (W9) at (-1.280,-1.530);
  \coordinate (B10) at (0.177,1.987);
  \coordinate (B11) at (0.652,0.868);
  \coordinate (B12) at (1.707,0.431);
  \coordinate (B13) at (-0.569,0.707);
  \coordinate (B14) at (-0.652,-0.868);
  \coordinate (B15) at (0.569,-0.707);
  \coordinate (B16) at (1.707,-1.374);
  \coordinate (B17) at (-0.177,-1.987);
  \coordinate (B18) at (-1.707,1.374);
  \coordinate (B19) at (-1.707,-0.431);
  \coordinate (d0) at (1.889,1.889);
  \coordinate (d1) at (0,2.67);
  \coordinate (d2) at (2.67,0);
  \coordinate (d3) at (1.889,-1.889);
  \coordinate (d4) at (0,-2.67);
  \coordinate (d5) at (-1.889,1.889);
  \coordinate (d6) at (-2.67,0);
  \coordinate (d7) at (-1.889,-1.889);

  \coordinate (C1) at (0.586,1.414);
  \coordinate (C2) at (1.160,0.743);
  \coordinate (C3) at (1.358,-0.563);
  \coordinate (C4) at (-0.295,1.346);
  \coordinate (C5) at (0,0);
  \coordinate (C6) at (0.295,-1.346);
  \coordinate (C7) at (-1.358,0.563);
  \coordinate (C8) at (-1.160,-0.743);
  \coordinate (C9) at (-0.586,-1.414);
  \path[facexfour,scale around={0.96:(C1)}] (1.0769,1.5579)--(0.3403,1.8631)
    arc[start angle=-37.20,end angle=-76.81,radius=0.205]--(0.2102,1.2719)
    arc[start angle=73.80,end angle=-7.84,radius=0.205]--(0.4888,0.9920)
    arc[start angle=142.78,end angle=61.20,radius=0.205]--(1.1058,1.4219)
    arc[start angle=-148.18,end angle=-187.81,radius=0.205]--cycle;
  \path[facexthree,scale around={0.96:(C2)}] (1.1813,1.3504)--(0.8262,0.9761)
    arc[start angle=31.82,end angle=-57.32,radius=0.205]--(0.8902,0.3027)
    arc[start angle=93.29,end angle=37.17,radius=0.205]--(1.5039,0.4033)
    arc[start angle=-172.22,end angle=-234.07,radius=0.205]--(1.3033,1.3263)
    arc[start angle=-83.46,end angle=-118.80,radius=0.205]--cycle;
  \path[facextwo,scale around={0.96:(C3)}] (0.8743,-0.1051)--(0.6928,-0.5436)
    arc[start angle=52.83,end angle=-52.83,radius=0.205]--(0.8743,-1.3089)
    arc[start angle=97.78,end angle=24.42,radius=0.205]--(1.5028,-1.3563)
    arc[start angle=175.03,end angle=82.17,radius=0.205]--(2.0550,-0.3992)
    arc[start angle=-127.22,end angle=-220.02,radius=0.205]--(1.7791,0.2391)
    arc[start angle=-69.41,end angle=-142.83,radius=0.205]--(1.1051,0.1257)
    arc[start angle=7.78,end angle=-97.78,radius=0.205]--cycle;
  \path[facexfive,scale around={0.96:(C4)}] (0.1062,1.2746)--(0.1198,1.7901)
    arc[start angle=-106.20,end angle=-141.55,radius=0.205]--(-0.6996,1.5443)
    arc[start angle=9.07,end angle=-52.83,radius=0.205]--(-0.5967,0.9101)
    arc[start angle=97.78,end angle=41.70,radius=0.205]--(-0.0473,1.0313)
    arc[start angle=-167.69,end angle=-256.81,radius=0.205]--cycle;
  \path[facexone,scale around={0.96:(C5)}] (-0.0001,0.9386)--(-0.3687,0.7507)
    arc[start angle=12.31,end angle=-97.78,radius=0.205]--(-0.7782,0.0654)
    arc[start angle=52.83,end angle=-57.32,radius=0.205]--(-0.6638,-0.6633)
    arc[start angle=93.29,end angle=-7.84,radius=0.205]--(-0.3162,-0.9510)
    arc[start angle=142.78,end angle=41.70,radius=0.205]--(0.3687,-0.7507)
    arc[start angle=-167.69,end angle=-277.78,radius=0.205]--(0.7782,-0.0654)
    arc[start angle=-127.17,end angle=-237.32,radius=0.205]--(0.6638,0.6633)
    arc[start angle=-86.71,end angle=-187.84,radius=0.205]--(0.3162,0.9510)
    arc[start angle=-37.22,end angle=-138.30,radius=0.205]--cycle;
  \path[facexfive,scale around={0.96:(C6)}] (-0.1062,-1.2746)--(-0.1198,-1.7901)
    arc[start angle=73.80,end angle=38.45,radius=0.205]--(0.6996,-1.5443)
    arc[start angle=-170.93,end angle=-232.83,radius=0.205]--(0.5967,-0.9101)
    arc[start angle=-82.22,end angle=-138.30,radius=0.205]--(0.0473,-1.0313)
    arc[start angle=12.31,end angle=-76.81,radius=0.205]--cycle;
  \path[facextwo,scale around={0.96:(C7)}] (-1.0887,1.4272)--(-1.5028,1.3563)
    arc[start angle=-4.97,end angle=-97.83,radius=0.205]--(-2.0550,0.3992)
    arc[start angle=52.78,end angle=-40.02,radius=0.205]--(-1.7791,-0.2391)
    arc[start angle=110.59,end angle=37.17,radius=0.205]--(-1.1051,-0.1257)
    arc[start angle=-172.22,end angle=-277.78,radius=0.205]--(-0.6928,0.5436)
    arc[start angle=-127.17,end angle=-232.83,radius=0.205]--(-0.8743,1.3089)
    arc[start angle=-82.22,end angle=-155.58,radius=0.205]--cycle;
  \path[facexthree,scale around={0.96:(C8)}] (-1.0654,-0.2218)--(-1.5039,-0.4033)
    arc[start angle=7.78,end angle=-54.07,radius=0.205]--(-1.3033,-1.3263)
    arc[start angle=96.54,end angle=61.20,radius=0.205]--(-0.8262,-0.9761)
    arc[start angle=-148.18,end angle=-237.32,radius=0.205]--(-0.8902,-0.3027)
    arc[start angle=-86.71,end angle=-142.83,radius=0.205]--cycle;
  \path[facexfour,scale around={0.96:(C9)}] (-0.3561,-1.0471)--(-0.4888,-0.9920)
    arc[start angle=-37.22,end angle=-118.80,radius=0.205]--(-1.1058,-1.4219)
    arc[start angle=31.82,end angle=-7.81,radius=0.205]--(-0.3403,-1.8631)
    arc[start angle=142.80,end angle=103.19,radius=0.205]--(-0.2102,-1.2719)
    arc[start angle=-106.20,end angle=-187.84,radius=0.205]--cycle;

  \foreach \a/\b in {
    W0/B10,W0/B11,W0/B12,W0/d0,
    W1/B10,W1/B13,W1/B11,
    W2/B11,W2/B15,W2/B12,
    W3/B12,W3/B16,W3/d2,
    W4/B14,W4/B17,W4/B15,
    W5/B15,W5/B17,W5/B16,
    W6/B10,W6/B18,W6/B13,
    W7/B13,W7/B19,W7/B14,
    W8/B18,W8/d6,W8/B19,
    W9/B14,W9/B19,W9/d7,W9/B17,
    B10/d1,B16/d3,B17/d4,B18/d5}
    \draw[plabicedge] (\a)--(\b);

  \foreach \n in {W0,W1,W2,W3,W4,W5,W6,W7,W8,W9}
    \node[plwhite] at (\n) {};
  \foreach \n in {B10,B11,B12,B13,B14,B15,B16,B17,B18,B19}
    \node[plblack] at (\n) {};

  \node[xlabel] at (0,0) {$x_1$};
  \node[xlabel] at (1.358,-0.563) {$x_2$};
  \node[xlabel] at (-1.358,0.563) {$x_2$};
  \node[xlabel] at (1.160,0.743) {$x_3$};
  \node[xlabel] at (-1.160,-0.743) {$x_3$};
  \node[xlabel] at (0.586,1.414) {$x_4$};
  \node[xlabel] at (-0.586,-1.414) {$x_4$};
  \node[xlabel] at (-0.295,1.346) {$x_5$};
  \node[xlabel] at (0.295,-1.346) {$x_5$};

  \node[facelabel] at (0,2.94) {$3$};
  \node[facelabel] at (2.079,2.079) {$2$};
  \node[facelabel] at (2.94,0) {$1$};
  \node[facelabel] at (2.079,-2.079) {$4$};
  \node[facelabel] at (0,-2.94) {$3$};
  \node[facelabel] at (-2.079,-2.079) {$2$};
  \node[facelabel] at (-2.94,0) {$1$};
  \node[facelabel] at (-2.079,2.079) {$4$};
  \end{scope}

  \draw[black,line width=1.55pt] (0,0) circle[radius=2.67];
\end{tikzpicture}
\end{minipage}%
\hfill
\begin{minipage}[c]{0.49\textwidth}
\centering
\begin{tikzpicture}[
  scale=1.02,
  plabicedge/.style={draw=black,line width=1.15pt,line cap=round},
  plblack/.style={circle,draw=black,line width=0.9pt,fill={rgb,255:red,143;green,137;blue,204},
    minimum size=3.0mm,inner sep=0pt},
  plwhite/.style={circle,draw=black,line width=0.95pt,fill=white,
    minimum size=3.0mm,inner sep=0pt},
  orientation/.style={draw=black,line width=0.82pt,
    -{Stealth[length=1.75mm,width=1.25mm]}},
  annularcut/.style={draw=red!85!orange,line width=1.00pt,
    dash pattern=on 4.0pt off 2.3pt,line cap=round,
    preaction={draw=white,line width=2.35pt,line cap=round}},
  faceback/.style={line width=0.90pt,line join=round,line cap=round},
  facexone/.style={faceback,draw={rgb,255:red,184;green,116;blue,129},
    fill={rgb,255:red,249;green,230;blue,233}},
  facextwo/.style={faceback,draw={rgb,255:red,102;green,143;blue,183},
    fill={rgb,255:red,226;green,238;blue,249}},
  facexthree/.style={faceback,draw={rgb,255:red,105;green,153;blue,118},
    fill={rgb,255:red,230;green,244;blue,232}},
  facexfour/.style={faceback,draw={rgb,255:red,191;green,137;blue,82},
    fill={rgb,255:red,250;green,237;blue,220}},
  facexfive/.style={faceback,draw={rgb,255:red,139;green,116;blue,177},
    fill={rgb,255:red,239;green,232;blue,248}},
  facelabel/.style={font=\small,inner sep=0.5pt},
  boundarylabel/.style={font=\small,inner sep=0pt},
  edgelabel/.style={font=\tiny,inner sep=0.35pt},
  externaledgelabel/.style={font=\tiny,inner sep=0.35pt}
]
  \begin{scope}[rotate=-45]
  \coordinate (W0) at (-0.544,2.030);
  \coordinate (W1) at (-0.92,0);
  \coordinate (W2) at (0.92,0);
  \coordinate (W3) at (2.030,-0.544);
  \coordinate (W4) at (-1.485,-1.485);
  \coordinate (B0) at (-2.030,0.544);
  \coordinate (B1) at (0,0.92);
  \coordinate (B2) at (1.485,1.485);
  \coordinate (B3) at (0,-0.92);
  \coordinate (B4) at (0.544,-2.030);
  \coordinate (dT) at (-0.691,2.579);
  \coordinate (dL) at (-2.579,0.691);
  \coordinate (dR) at (2.579,-0.691);
  \coordinate (dB) at (0.691,-2.579);

  \coordinate (C1q) at (0,0);
  \coordinate (C2q) at (0.582,-0.582);
  \coordinate (C3q) at (0.95,0.95);
  \coordinate (C4q) at (-0.98,0.98);
  \coordinate (C5q) at (-0.95,-0.95);
  \path[facexone,even odd rule] (-0.7280,0.0718)
    --(-0.0718,0.7280) arc[start angle=-110.50,end angle=-69.50,radius=0.205]
    --(0.7280,0.0718) arc[start angle=159.50,end angle=200.50,radius=0.205]
    --(0.0718,-0.7280) arc[start angle=69.50,end angle=110.50,radius=0.205]
    --(-0.7280,-0.0718) arc[start angle=-20.50,end angle=20.50,radius=0.205]--cycle
    (0,0) circle[radius=0.255];
  \path[facextwo] (0.1920,-0.8482)
    --(0.8482,-0.1920) arc[start angle=-110.50,end angle=44.67,radius=0.205]
    --(1.4710,1.2093)
    --(1.8995,-0.3859) arc[start angle=129.53,end angle=200.50,radius=0.205]
    --(0.6158,-1.8380) arc[start angle=69.50,end angle=140.47,radius=0.205]
    --(-1.2093,-1.4710)
    --(-0.1441,-1.0658) arc[start angle=-134.67,end angle=20.50,radius=0.205]--cycle;
  \path[facexthree] (-0.3856,1.8999)
    --(-0.0058,1.1249) arc[start angle=91.61,end angle=-20.50,radius=0.205]
    --(0.8482,0.1920) arc[start angle=110.50,end angle=93.67,radius=0.205]
    --(1.3392,1.3409) arc[start angle=-135.33,end angle=-170.54,radius=0.205]
    --(-0.3859,1.8995) arc[start angle=-39.53,end angle=-39.39,radius=0.205]--cycle;
  \path[facexfour] (-0.6158,1.8380)
    --(-1.8380,0.6158) arc[start angle=20.50,end angle=-1.61,radius=0.205]
    --(-1.0501,0.1584) arc[start angle=129.39,end angle=69.50,radius=0.205]
    --(-0.1920,0.8482) arc[start angle=-159.50,end angle=-219.39,radius=0.205]
    --(-0.5382,1.8251) arc[start angle=-88.39,end angle=-110.50,radius=0.205]--cycle;
  \path[facexfive] (-1.8999,0.3856)
    --(-1.1249,0.0058) arc[start angle=178.39,end angle=290.50,radius=0.205]
    --(-0.1920,-0.8482) arc[start angle=159.50,end angle=176.33,radius=0.205]
    --(-1.3409,-1.3392) arc[start angle=45.33,end angle=80.54,radius=0.205]
    --(-1.8995,0.3859) arc[start angle=-50.47,end angle=-50.61,radius=0.205]--cycle;

  \foreach \a/\b in {
    W0/B0,W0/B1,W0/B2,W0/dT,
    W1/B0,W1/B1,W1/B3,
    W2/B1,W2/B2,W2/B3,
    W3/B2,W3/B4,W3/dR,
    W4/B0,W4/B3,W4/B4,
    B0/dL,B4/dB}
    \draw[plabicedge] (\a)--(\b);

  \foreach \a/\b in {
    W0/B0,W0/B1,W0/B2,dT/W0,
    W1/B0,B1/W1,W1/B3,
    W2/B1,B2/W2,W2/B3,
    W3/B2,dR/W3,
    W4/B0,B3/W4,W4/B4,
    B0/dL,B4/dB}
    \draw[orientation] ($(\a)!0.39!(\b)$)--($(\a)!0.61!(\b)$);

  \draw[orientation] ($(W3)!0.20!(B4)$)--($(W3)!0.43!(B4)$);

  \foreach \n in {W0,W1,W2,W3,W4}
    \node[plwhite] at (\n) {};
  \foreach \n in {B0,B1,B2,B3,B4}
    \node[plblack] at (\n) {};

  \node[facelabel] at (0.46,-0.106) {$x_1$};
  \node[facelabel] at (1.10,-0.78) {$x_2$};
  \node[facelabel] at (0.693,1.047) {$x_3$};
  \node[facelabel] at (-0.860,0.860) {$x_4$};
  \node[facelabel] at (-1.047,-0.693) {$x_5$};

  \node[boundarylabel] at (-0.745,2.780) {$2$};
  \node[boundarylabel] at (-2.780,0.745) {$3$};
  \node[boundarylabel] at (2.780,-0.745) {$1$};
  \node[boundarylabel] at (0.745,-2.780) {$4$};
  \end{scope}

  \path (W0)--(B0) node[edgelabel,pos=0.50,sloped,above=1.8pt] {$\alpha_1$};
  \coordinate (alphatwomid) at ($(W0)!0.50!(B1)$);
  \node[font=\tiny,inner sep=0.35pt,rotate=67.5]
    at ([xshift=5.5pt,yshift=-2.0pt]alphatwomid) {$\alpha_2$};
  \path (W0)--(B2) node[edgelabel,pos=0.50,sloped,above=1.8pt] {$\alpha_3$};
  \path (W0)--(dT) node[externaledgelabel,pos=0.72,left=2.0pt] {$\alpha_4$};
  \coordinate (alphafivemid) at ($(W1)!0.50!(B0)$);
  \node[font=\tiny,inner sep=0.35pt,rotate=-67.5]
    at ([xshift=-5.5pt,yshift=-2.0pt]alphafivemid) {$\alpha_5$};
  \coordinate (alphasixmid) at ($(W1)!0.50!(B3)$);
  \node[edgelabel,rotate=90]
    at ([xshift=-5.3pt]alphasixmid) {$\alpha_6$};
  \coordinate (alphasevenmid) at ($(W1)!0.50!(B1)$);
  \node[font=\tiny,inner sep=0.35pt]
    at ([yshift=6.8pt]alphasevenmid) {$\alpha_7$};
  \coordinate (alphaeightmid) at ($(W2)!0.50!(B1)$);
  \node[edgelabel,rotate=90]
    at ([xshift=5.3pt]alphaeightmid) {$\alpha_8$};
  \coordinate (alphaninemid) at ($(W2)!0.50!(B3)$);
  \node[edgelabel]
    at ([xshift=1.0pt,yshift=-5.5pt]alphaninemid) {$\alpha_9$};
  \coordinate (alphatenmid) at ($(W2)!0.50!(B2)$);
  \node[edgelabel,rotate=22.5]
    at ([xshift=-4.0pt,yshift=-8.0pt]alphatenmid) {$\alpha_{10}$};
  \coordinate (alphaelevenmid) at ($(W3)!0.50!(B2)$);
  \node[edgelabel,rotate=67.5]
    at ([xshift=5.0pt,yshift=-2.1pt]alphaelevenmid) {$\alpha_{11}$};
  \path (W3)--(B4) node[edgelabel,pos=0.35,sloped,below=1.8pt] {$\alpha_{12}$};
  \path (W3)--(dR) node[externaledgelabel,pos=0.72,left=2.0pt] {$\alpha_{13}$};
  \coordinate (alphafourteenmid) at ($(W4)!0.50!(B3)$);
  \node[edgelabel,rotate=-22.5]
    at ([xshift=4.0pt,yshift=-8.0pt]alphafourteenmid) {$\alpha_{14}$};
  \path (W4)--(B0) node[edgelabel,pos=0.50,sloped,above=1.8pt] {$\alpha_{15}$};
  \coordinate (alphasixteenmid) at ($(W4)!0.50!(B4)$);
  \node[edgelabel,rotate=-67.5]
    at ([xshift=-5.0pt,yshift=-2.1pt]alphasixteenmid) {$\alpha_{16}$};
  \path (B0)--(dL) node[externaledgelabel,pos=0.72,right=2.0pt] {$\alpha_{17}$};
  \path (B4)--(dB) node[externaledgelabel,pos=0.72,right=2.0pt] {$\alpha_{18}$};

  \begin{scope}[rotate=-45]
    \coordinate (cutend) at (1.531,-2.187);
    \draw[annularcut] (0,0)
      .. controls (0.248,-0.389) and (0.233,-0.969) .. (0.707,-1.131)
      .. controls (1.181,-1.294) and (1.379,-1.732) .. (cutend);
    \node[font=\scriptsize,text=red!85!orange,inner sep=0.5pt]
      at (0.61,-1.30) {$\lambda$};
  \end{scope}

  \fill[white] (0,0) circle[radius=0.17];
  \draw[black,line width=1.55pt] (0,0) circle[radius=0.17];
  \draw[black,line width=1.55pt] (0,0) circle[radius=2.67];
\end{tikzpicture}
\end{minipage}
\caption{Network realization of the positive $\mathcal X$-coordinate
parametrization.  Left: the
disk network associated with the symmetric $\Gr(4,8)$ quiver; faces related
by the folding carry the same color and the same variable.  Right: its
annular quotient, with the induced perfect orientation.  The red dashed cut
records winding around the puncture, and a path crossing it acquires a factor
of $\lambda$.}
\label{fig:periodic-bipartite-networks}
\end{figure}
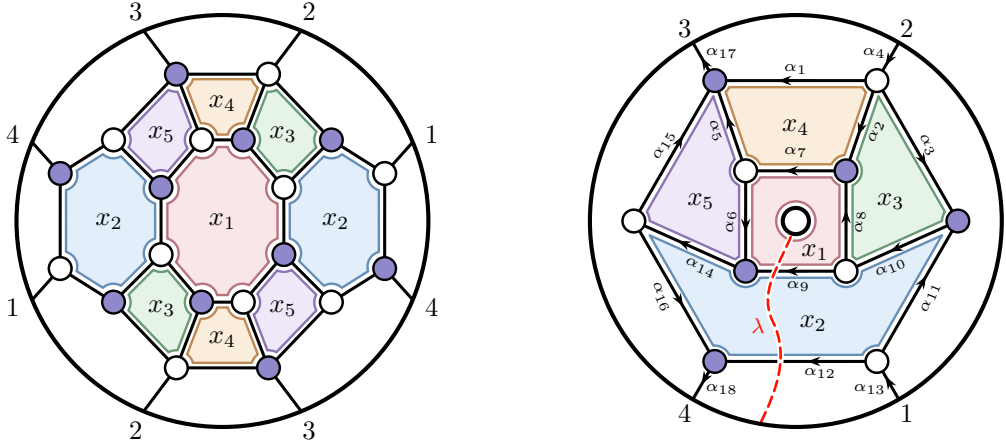